\documentclass[12pt]{article}
\usepackage{graphicx,amsmath}
\usepackage{units}
\usepackage[usenames]{color}

\newlength{\dinwidth}
\newlength{\dinmargin}
\renewcommand{\vec}[1]{\boldsymbol{#1}}

\def\lapproxeq{\lower .7ex\hbox{$\;\stackrel{\textstyle                                                    
<}{\sim}\;$}}                                                    
\def\gapproxeq{\lower .7ex\hbox{$\;\stackrel{\textstyle                                                    
>}{\sim}\;$}}                                                    
\def\be{\begin{equation}}                                                    
\def\ee{\end{equation}}                                                    
\def\bea{\begin{eqnarray}}                                                    
\def\eea{\end{eqnarray}}
\def\b{\vec{b}}
     
\def\q{\vec{q}}

\def\GeV{\rm GeV}

\def\sh{\hat s}
\def\sh2{{\hat s}^2}

\def\el{\sigma_{\rm el}}
\def\DD{\sigma_{\rm DD}}
\def\SD{\sigma_{\rm SD}}

\begin{document}

\titlepage                                                    
\begin{flushright}                                                    
IPPP/14/14  \\
DCPT/14/28 \textbf{}\\                                                    
\today \\                                                    
\end{flushright} 
\vspace*{0.5cm}
\begin{center}                                                    
{\Large \bf Elastic scattering and Diffractive dissociation}\\
\vspace*{0.5cm}
{\Large \bf in the light of LHC data}\footnote{Contribution to the special issue of the International Journal of Modern Physics A on "Elastic and diffractive scattering" coordinated by Christophe Royon.}\\
\vspace*{1cm}
 V.A. Khoze$^{a,b}$, A.D. Martin$^a$ and M.G. Ryskin$^{a,b}$\\                                                 
\vspace*{0.5cm}                                                    
$^a$ Institute for Particle Physics Phenomenology, University of Durham, Durham, DH1 3LE \\                                                   
$^b$ Petersburg Nuclear Physics Institute, NRC Kurchatov Institute, Gatchina, St.~Petersburg, 188300, Russia

\vspace*{1cm}    
\begin{abstract}

We study the behaviour of elastic and diffractive proton dissociation cross sections at high energy. First, we describe what would be expected to be observed at the LHC based on conventional Regge theory. We emphasize the tension between these expectations and the recent LHC measurements, and we discuss the possibilty to modify the classic Reggeon Field Theory in a physically-motivated way so as to accommodate the tendencies observed at the LHC.
As a result, we show that we are able to achieve a  `global' description of the wide variety of high energy elastic and diffractive data that are presently available, particularly from the LHC experiments. The model is based on only one pomeron pole, but includes multi-pomeron interactions and, {\it significantly}, includes the transverse momentum dependence of intermediate partons as a function of their rapidity, which provides the rapidity dependence of the multi-pomeron vertices.
We give predictions for diffractive observables at LHC, and higher, energies.

\end{abstract}                                                        
\vspace*{0.5cm}                                                    
                                                    
\end{center}

\section{Introduction}
Elastic proton-proton scattering is an important fundamental reaction, which provides information not only about the $2\to 2$ strong interaction amplitude, but, via unitarity, it contains generalized (integrated) information
about the sum of all inelastic processes as well.

The experimental and theoretical study of elastic 
interactions has a long history (see e.g \cite{abk,Goulianos:1982vk,Block:2006hy,Fiore:2008tp,KMRJphysG,
Dremin:2012ke}). 
Rather than presenting a complete review, here we concentrate on some non-trivial tendencies of the high energy amplitude which can be seen from the LHC Run 1 data. 
Processes, such as $pp\to X+p$ and $pp\to X+Y$, where one or both protons dissociate, are intimately connected to elastic scattering $pp\to p+p$. Here the + sign indicates the presence of a large rapidity gap. We survey the experimental information available for such high-energy diffractive processes, and discuss how to obtain a global overall framework to simultaneously describe all these data.
We emphasize the phenomenological features extracted from such a global description of these measurements, in which the recent LHC data play a major role.

 We start by discussing the description of high-energy elastic scattering based on pomeron exchange, and on how unitarity is preserved using an eikonal approach. In Section \ref{sec:extend} we extend the framework to allow for quasi-elastic processes, such as $pp\to p+N^*$.
In Section \ref{sec:4}  we show how to formulate the description of proton dissociation into high mass states, which involves the introduction of multi-pomeron couplings.   In Section \ref{sec:data} we consider the experimental data for high-energy diffractive processes -- and emphasize the puzzles one faces in trying to understand the qualitative behaviour of the recent LHC measurements, which considerably extend the types of measurements that have become available. Section \ref{sec:global} describes an attempt to obtain a `global' simultaneous description of all these various measurements. In Section \ref{sec:6} we present our conclusions, and  predictions for diffractive observables at various energies.  We recall some of the main properties of the QCD pomeron in Appendices A and B.  

     We shall see that, in order to reproduce the qualitative features of the LHC diffractive data, we are led to introduce the energy and rapidity dependence of the proton-pomeron and multi-pomeron couplings. From the old Reggeon point of view this dependence may seem unexpected. Recall, however, that these couplings are dimensionful quantities, whose values are controlled by the corresponding transverse momenta of partons. On the other hand, the experiments at the LHC have already observed the growth of the $k_t$ of secondaries with increasing collider energy. This growth is a natural consequence of the diffusion in ln$k_t$ space predicted for the QCD pomeron within the BFKL approach. In this paper we simply parametrise the $k_t$ behaviour to match,  simultaneously, the wide variety of available high-energy diffractive data obtained from the LHC experiments.                                                
The origin of the low $k_t$ diffusion and the role of the absorptive effects are discussed in Appendix A.

\section{Eikonal approach to elastic scattering   \label{sec:2}} 

   Diffractive processes caused by pomeron exchange are usually described 
within the framework of Reggeon Field Theory (RFT)~\cite{Gribov}.
In the simplest case, the high energy elastic scattering amplitude, $T_{\rm el}$, (and correspondingly the total cross section) is parametrized by single pomeron exchange. The trajectory of this effective 
(soft) pomeron reads
\be
\label{eq:pom-tr}
\alpha_P(t)=1+\Delta +\alpha'_P t\ ,
\ee 
with $\Delta=0.08$ and $\alpha'_P=0.25$ GeV$^{-2}$~\cite{DL}.

However, already two-particle $s$-channel unitarity generates a series of the 
non-enhanced multi-pomeron diagrams leading to the eikonal approximation.   The unitarity relation is of the form
\be
2~{\rm Im}\ T_{\rm el}(b)= |T_{\rm el}(b)|^2+G_{\rm inel}(b)\ ,
\ee
where $G$ is the sum over all the inelastic intermediate states.
The solution gives an elastic amplitude,
\be
\label{eq:1i}
T_{\rm el}~=~i(1-e^{-\Omega/2})\ ,
\ee
where one pomeron exchange describes the opacity $\Omega(s,b)$, which depends on the square of the incoming energy, $s$, and the impact parameter, $b$. The opacity plays the role of the phase, $i\delta_l$, of the amplitude with orbital momentum $l=b\sqrt s/2$; that is $\Omega/2=2i\delta_l$.
\bea 
\sigma_{\rm tot}(s,b) & = & 2(1-{\rm e}^{-\Omega/2}), \label{eq:tot}\\
\sigma_{\rm el}(s,b) & = & (1-{\rm e}^{-\Omega/2})^2, \label{eq:el}\\
\sigma_{\rm inel}(s,b) & = & 1- {\rm e}^{-\Omega}, \label{eq:inel}
 \eea
 Note, from (\ref{eq:inel}), that
 \be 
 S^2(b)\equiv e^{-\Omega}
 \label{eq:S2}
 \ee
  is the probability that  no inelastic interaction occurs at impact parameter $b$. Later, we will see that this observation will enable us to calculate the probability that large rapidity gaps survive soft rescattering.

 In terms of the opacity the elastic cross section takes the form
\be
\frac{d\sigma_{\rm el}}{dt}=\frac{1}{4\pi}  \left| \int d^2b~e^{i\q_t \cdot \b} (1-e^{-\Omega(b)/2}) \right|^2=\frac{1}{2}  \left| \int bdb~J_0(q_tb) (1-e^{-\Omega(b)/2}) \right|^2
\label{eq:el6}
\ee
where $q_t=\sqrt{|t|}$.

\subsection{Elastic cross section at fixed $s$}
As far as we are going to describe the elastic scattering at one fixed energy it is not a problem to find an appropriate parametrization for the opacity $\Omega(b)$ and to tune the parameters to reproduce the observed $d\sigma_{\rm el}/dt$ cross section.
Moreover, one can fix the form of the parametrization, but choose, at each particular energy, the corresponding values of parameters; see, for example, \cite{FGPS}
and \cite{UG,Uzhinsky:2011qu}.
\begin{figure} 
\begin{center}
\vspace{-6.0cm}
\includegraphics[height=15cm]{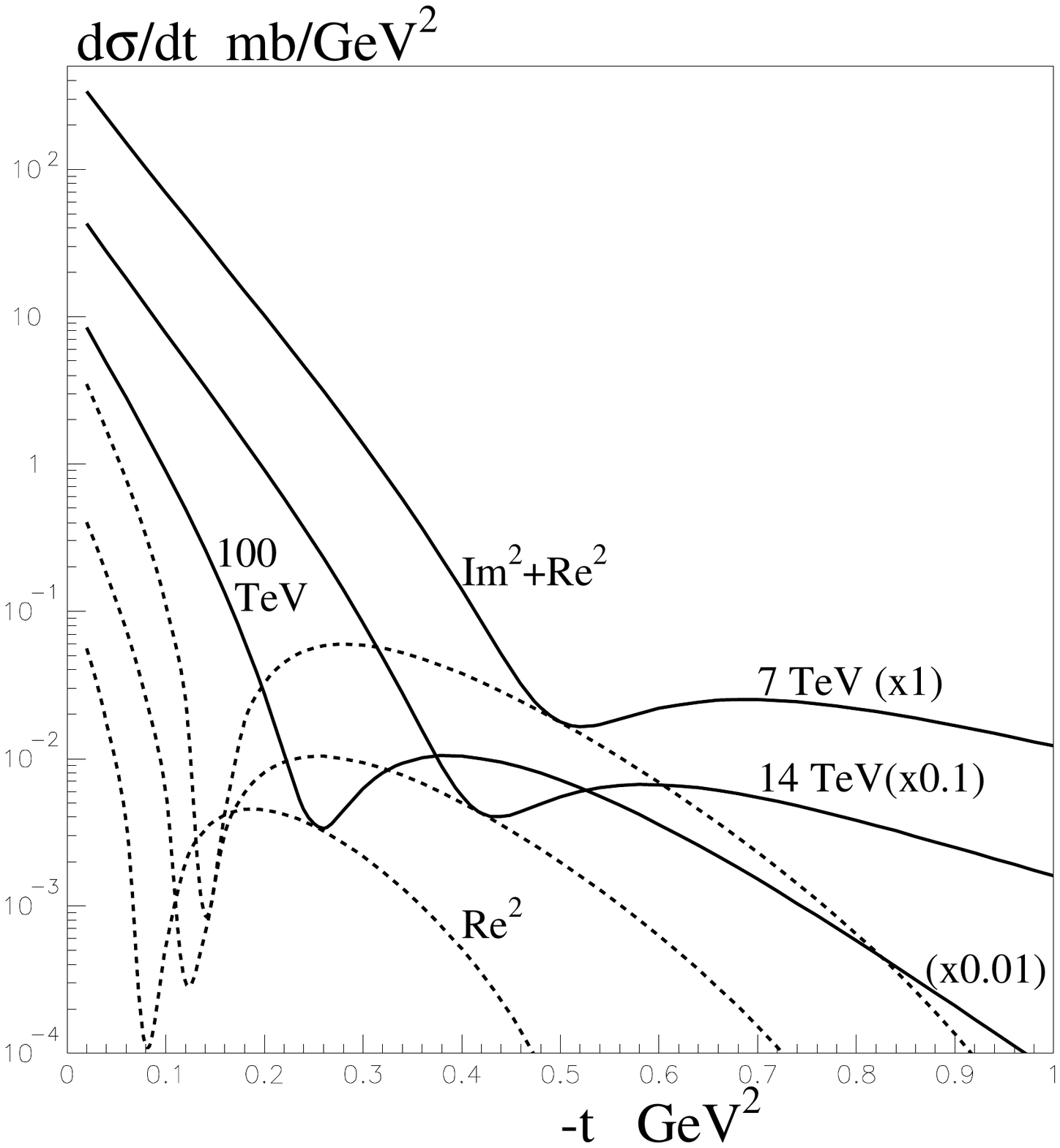}
\vspace{-0.2cm}
\caption{\sf The continuous curves are the total contribution to the elastic cross section, and the dashed curves are the contribution of the real part of the amplitude, at three collider energies: 7, 14 and 100 TeV. }
\label{fig:re-im}
\end{center}
\end{figure}

 Alternatively, we may simply take the Fourier transform from the experimental
 data~\cite{Amaldi,Amaldi:1976gr}
\begin{equation}
{\rm Im}T(b)~=~\int \sqrt{\frac{d\sigma_{\rm el}}{dt}\frac{16\pi}{1+\rho^2}}~ J_0(q_tb)~ \frac{q_tdq_t}{4\pi},
\label{eq:2i}
\end{equation}
where square root represents Im$T_{\rm el}(q_t),$ with $\rho \equiv {\rm Re}T/{\rm Im}T$. In this way, we first determine Im$T$ from the data for $d\sigma_{\rm el}/dt$, and then calculate $\Omega (b)$ using {(\ref{eq:1i}), assuming that $\rho$ is small  or   $\rho(t)=$constant. In fact, later on, we assume $\rho^2\ll 1$ (which is actually well justified except in the diffractive dip region; see Fig. \ref{fig:re-im}, where we present the $\rho^2$-contribution to the elastic cross section, $d\sigma_{\rm el}/dt$, calculated within the model described in Section \ref{sec:global}} at $\sqrt{s}=7,~14$ and 100 TeV).
In other words the elastic cross section probes the optical density (i.e. 
the opacity) of the proton.

\begin{figure} 
\begin{center}
\vspace{-6.0cm}
\includegraphics[height=15cm]{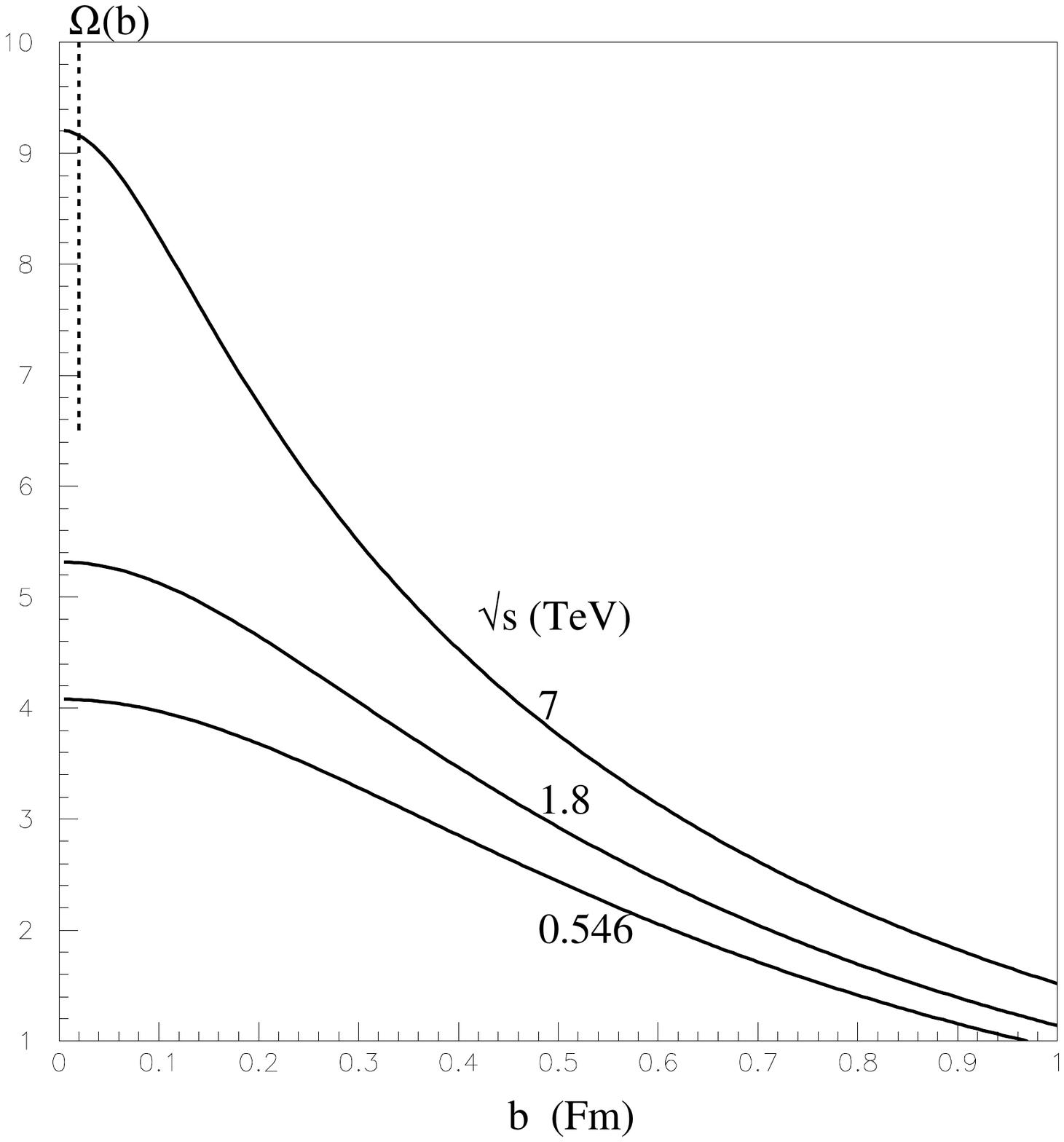}
\vspace{-0.2cm}
\caption{\sf The proton opacity $\Omega(b)$ determined directly from the  $d\sigma_{\rm el}/dt$ data at 546 GeV \cite{SppS,Augier:1993sz,Arnison:1983mm}, 
1.8 TeV \cite{EEE,cdfB} and 7 TeV \cite{TOTEM2} data.  The uncertainty on the LHC value at $b=0$ is indicated by a dashed line.}
\label{fig:tt1}
\end{center}
\end{figure}

 The results obtained via (\ref{eq:2i}) are shown in Fig.~\ref{fig:tt1}, where 
we compare $\Omega(b)$ obtained from elastic differential cross section data at S$p\bar{p}$S 
\cite{SppS,Augier:1993sz,Arnison:1983mm}, 
Tevatron \cite{EEE,cdfB} and LHC \cite{TOTEM2} energies. At the lower two energies the $\Omega (b)$ distributions have approximately Gaussian form, whereas at the LHC energy we observe a growth of $\Omega$ at small $b$. The growth reflects the fact that the TOTEM data indicate that we have almost total saturation at $b=0$. Note that, according to (\ref{eq:1i}), the
 value of Im$T(b=0)\to 1$ corresponds to $\Omega \to \infty$. 
Since actually we do not reach  exact saturation, the proton opacity at $b=0$ is not $\infty$, but just large numerically.  Clearly, in this region of $b$, the uncertainty on the value of $\Omega$ is large as well, see Fig.~\ref{fig:tt1}.   

\subsection{Elastic scattering as a function of both $s$ and $t$}
 It is more complicated to describe simultaneously both the $t$ and the 
energy ($\sqrt{s}$) dependence of the elastic cross sections. That is to describe the energy dependence of the opacities, $\Omega(b)$, obtained via (\ref{eq:2i}) or in the previous (fixed energy) analysis.

We first discuss the high-energy behaviour expected in the eikonal model (\ref{eq:1i}) with the opacity given by the exchange of one pomeron with a linear trajectory of slope $\alpha'_P$ and intercept $\alpha_P(0)>1$. In this case the corresponding opacity is
  \be
\label{eq:oob}
\Omega(s,b)=\int\frac{d^2q_t}{4\pi^2}~\Omega(s,q_t)~e^{i\q_t \cdot \b}
\ee
with
\be
\Omega(s,q_t)=-i\eta_P(t) g_N(t) g_N(t) \left(\frac s{s_0}\right)^{\alpha_P(t)-1}
\ ,
\label{eq:oot}
\ee
where  $g_N(t)$ is the proton-pomeron coupling and where the conventional dimensionful scale $s_0=1$ GeV$^2$ is taken; finally $\eta_P$ is the signature factor of the pomeron
\be
\eta_P(t)~=~-\frac{1+{\rm exp}(-i\pi\alpha_P(t))}{{\rm sin}\pi\alpha_P(t)}.
\label{eq:signature}
\ee

If we assume, as usual, an exponential $t$ dependence of the coupling, $g_N(t)=g_N(0)\exp(b_0t)$, then the opacity generated by one pomeron pole is 
\be
{\rm Re}~\Omega(s,q_t)= g_N(0) g_N(0) \left(\frac s{s_0}\right)^{\alpha_P(0)-1} {\rm exp}(Bt)
\label{eq:B}
\ee
where the $t$-slope of the amplitude is
\be
B~=~2b_0+\alpha_P'~{\rm ln}\left(\frac{s}{s_0}\right).
\label{eq:slope}
\ee
We consider here just the real part of the opacity $\Omega$, neglecting the phase of pomeron exchange. For more precise calculations we restore the real part of the amplitude $T(b)$ by making use of dispersion relations, see Section \ref{sec:elas}.  
Inserting (\ref{eq:B}) into (\ref{eq:oob}) we find that the high-energy amplitude has a gaussian form in $b$-space
\be
\label{eq:exp}
{\rm Re}~\Omega(s,b)=\frac{g^2_N(0)}{4\pi B}\left(\frac s{s_0}\right)^{\alpha_P(0)-1}
\exp(-b^2/4B)\ ,
\ee

So, to reiterate, the exchange of a {\it single} pomeron pole gives a total cross section with a power-like energy dependence,
\be
 \sigma_{\rm tot}~\propto~s^\Delta~~~~~~~{\rm with}~~~~~~\Delta\equiv\alpha_P(0)-1,
\ee
and an amplitude with a Gaussian profile in $b$-space, with an effective radius of interaction which increases at high energies as $\sqrt{\alpha'_P {\rm ln}(s/s_0)}$. If we take for the pomeron the DL parametrisation \cite{DL}, then, at  LHC energies,  the Gaussian exceeds the black disc limit for small $b$.

However, eikonal unitarization damps the power growth of the one pomeron exchange cross section. Thus, in (\ref{eq:exp}), $\Omega(s,b)\propto (s/s_0)^{\alpha_P-1}$ gives an amplitude Im$\ T_{\rm el}(s,b)=1-e^{-\Omega/2}<1$. Hence the 
total cross section is limited by the size of the effective interaction area $\sigma_{\rm tot}<2\pi R^2$, where the interaction radius $R$ can be estimated from (\ref{eq:exp}). To be explicit, at high energies we may write (\ref{eq:exp}) as
\bea
\label{eq:exp2}
\frac{\Omega(s,b)}{2}~& = &~  \frac{g^2_N(0)}{8\pi B}{\rm exp}\left(\Delta{\rm ln} (s/s_0)-\frac{b^2}{4\alpha'~{\rm ln(s/s_0)}}\right),\\
 &\gg &~1~~~~~~~~~~{\rm for}~~b^2<R^2=4\Delta\alpha'~{\rm ln}^2(s/s_0)
\eea
and so tends to the black disc limit  for $b \lapproxeq R$.
Thus we expect that at higher energies the cross section will increase slower than the $s^{0.08}$ predicted by the Donnachie-Landshoff fit~\cite{DL}.

The next observation is that we expect the $t$-slope $B$ of the elastic cross section to increase with $|t|$ up to the first diffractive dip. Indeed, let us start with a relatively small one-pomeron 
amplitude and consider the two-pomeron contribution corresponding to the $\Omega^2$  term in the expansion of the eikonal  $1-\exp(-\Omega/2)$. In this term the momentum transferred,
$q_t=\sqrt{|t|}$, is divided between the two pomerons so that each pomeron carries about a momentum $q_t/2$. Correspondingly the $t$ dependence of the whole 'two-pomeron' amplitude will be $\exp(2B(t/4))=\exp(Bt/2)$. In other words, the two pomeron contribution, which has an opposite sign in comparison with the one pomeron exchange, has a twice smaller $t$-slope\footnote{In terms of the impact parameter representation, the $\Omega^2(b)$ term is concentrated in a domain of a smaller radius.}. Therefore the effective slope of the whole amplitude increases when the higher $\Omega^n$ order terms start to cancel the one-pomeron contribution~\footnote{At very high energies, the amplitude $T_{el}(b)$ takes the form of the $\Theta$ function, $T_{el}(b)\simeq i\Theta(R(s)-b)$. The corresponding Fourier integral (\ref{eq:el6}) gives the Bessel function $J_1(\sqrt tR)/\sqrt tR$, which has a $t$-slope which increases with $|t|$.} and this effect should be more pronounced at a higher energies when the diffractive dip occurs at  lower $|t|$ values.

\subsection{Surprises in the LHC elastic data   \label{sec:surprise}} 
 
 Prior to the LHC (in the pre-LHC era) the energy behaviour of elastic $pp,~p\bar{p},~\pi p,~ Kp$ cross 
sections $d\sigma_{\rm el}/dt$ (i.e. the $\Omega(b,s)$) was satisfactorily reproduced by the sum of the pomeron and the secondary Reggeon contributions (see, for example,~\cite{BS}.
However above Tevatron energies the secondary Reggeons contributions completely die out and the situation illuminates the properties of the pomeron alone. Contrary to the expectations described above, the total cross section in the Tevatron -- LHC energy interval starts to grow {\it faster}, not more slowly, than below the Tevatron energy, and, secondly, the slope of the effective pomeron trajectory, $\alpha'_{\rm eff}$, increases \cite{Schegelsky}. 

In particular, 
the DL fit predicts $\sigma_{\rm tot}=90.7$ mb at $\sqrt{s}=7$ TeV, while TOTEM
observes 98.6$\pm 2.2$ mb \cite{TO}. Moreover, the elastic slope was measured
at the Tevatron ($\sqrt s=1.8$ TeV) to be $B_{\rm el}=16.3\pm 0.3$ GeV$^{-2}$ by
the E710  experiment \cite{EEE} and to be
$B_{\rm el}=16.98\pm 0.25$ GeV$^{-2}$ by the CDF group \cite{cdfB}. Even starting from
the CDF result, and using the $\alpha'_P=0.25$ GeV$^{-2}$, we expect, from (\ref{eq:slope}),
 \be 
 B_{\rm el}=16.98+4\times 0.25\times \ln(7/1.8)=18.34~ \GeV^{-2}
 \ee
  at 7 TeV, while TOTEM finds $19.9\pm 0.3$ GeV$^{-2}$ \cite{TO}.
  Next, in the relatively low $|t|<0.3\ -\ 0.4$ GeV$^2$ region
we do not see the expected increase of $t$-slope as $|t|$ increases.


\section{Extending the eikonal approach   \label{sec:extend}}
The one-channel eikonal described in Section \ref{sec:2} is a rather naive oversimplified approximation. It provides hints of the behaviour
we may expect of elastic cross section, but clearly it does not give the whole story.
 Moreover, even within the framework of the one-channel eikonal, the
expectation that $t$-slope grows with $|t|$ may be masked by other 
effects. First, there is no reason why the $t$ dependence of the proton-pomeron coupling $g_N(t)$ has to be a pure exponent. If, instead, we were to assume a power dependence, like $g_N(t)\propto 1/(1-t/t_0)^2$, then the effective slope generated by the coupling {\em decreases} with $|t|$. Next, there exists a two-pion singularity at $t=4m^2_\pi$ (close to the physical region) in the pomeron trajectory (\ref{eq:pom-tr}) which also generates some {\em positive} curvature (that is, some decrease of $t$-slope) in the behaviour of $d\sigma_{\rm el}/dt$~\cite{AG,KMR18}. Moreover, this last effect also increases with energy. So there may be some compensation between the {\it negative} curvature caused by the eikonal (arising from the interference between the different multi-pomeron contributions), and the {\it positive} curvature coming from the form of the proton-pomeron coupling and the two-pion singularity of the pomeron trajectory. However, an exact compensation looks quite non-trivial and it is not easy to explain the observed
almost {\em pure} exponential behaviour of $d\sigma_{\rm el}/dt$ up to $|t|\sim 0.3\ -\ 0.4$ GeV$^2$. Of course, it is possible to reproduce such a behaviour by tuning parameters, but the reasons for this "fine tuning" are not known. At the moment, it appears to be just an accident.

\subsection{Non-locality of the proton (and of the pomeron)}

In order to make the analysis more realistic and self-consistent we have to discuss, not only the elastic process, but the whole set of the soft phenomena, including the diffractive dissociation of the incoming protons -- that is the single and double proton dissociation processes $pp\to X+p$ and $pp\to X+Y$ where the + sign denotes the presence of a large rapidity gap.  Recently, data from Run 1 of the LHC have become available for these processes, and force us to study the phenomenological description of soft phenomena in more detail.      The observations of proton dissociation indicate that we  have to take account of the fact that neither the pomeron nor the proton are the local elementary objects but have their own structure. As we will see, there are sound theoretical reasons why we must include this structure in a description of the data for `soft' high energy interactions.

Indeed, 
diffractive dissociation can {\it only} occur as a consequence of the {\em internal
structure} of the proton. This is simplest to describe at high
energies, where the lifetimes of the hadronic fluctuations are
large, $\tau\sim E/m^2$, and during these time intervals the
corresponding Fock states can be considered as `frozen'. Each
hadronic constituent can undergo scattering and thus destroy the
coherence of the fluctuations. As a consequence, the outgoing
superposition of states will be different from the incident
proton, and will most likely contain multiparticle states, so we
will have diffractive dissociation, as well as elastic scattering.

We will discuss the structure of the pomeron in Section  \ref{sec:4}, but, first, we introduce the Good-Walker formalism for (low-mass) diffractive dissociation.

\subsection{Good-Walker approach: low-mass dissociation  \label{sec:GW}}
To discuss diffractive dissociation, it is convenient to follow Good
and Walker~\cite{GW}, and to introduce states $\phi_k$ which
diagonalize the $T$ matrix which describes different $p\to N^*,~ N^*_a\to N^*_b$ transitions caused by the nucleon-pomeron couplings, $g_{ab}$. Such eigenstates only undergo elastic
scattering. Since there are no off-diagonal transitions
\be \langle \phi_j|T|\phi_k\rangle = 0\qquad{\rm for}\ j\neq k \ee
a state $k$ cannot diffractively dissociate in a state $j$. We
 have noted that this is not true for the proton due to its
internal structure, So one way of proceeding is to enlarge the set of
intermediate states $(p,N^*_a)$, from just the single elastic channel, and to
introduce a multichannel eikonal.  Another possibility is to work in terms of the Good-Walker eigenstates $\phi_j$, having a simple one-channel eikonal for each state.

Let us express the cross section in terms of the
probabilities $F_k$ of the hadronic process proceeding via the
various diffractive eigenstates $\phi_k$.
We denote the orthogonal matrix which diagonalizes ${\rm
Im}\,T$ by $a$, so that
\be \label{eq:a3} {\rm Im}\,T \; = \; aFa^T \quad\quad {\rm with}
\quad\quad \langle \phi_j |F| \phi_k \rangle \; = \; F_k \:
\delta_{jk}. \ee
Now consider the diffractive dissociation of an incoming proton described by the state $|i\rangle$. We may write
\be \label{eq:a4} | i \rangle \; = \; \sum_k \: a_{ik} \: | \phi_k
\rangle. \ee
The elastic scattering amplitude satisfies
\be \label{eq:a5} \langle i |{\rm Im}~T| i \rangle \; = \; \sum_k
\: |a_{ik}|^2 \: F_k \; = \; \langle F \rangle, \ee
where $F_k \equiv \langle \phi_k |F| \phi_k \rangle$ and where the
brackets of $\langle F \rangle$ mean that we take the average of
$F$ over the initial probability distribution of diffractive
eigenstates. After the diffractive scattering described by
$T_{fi}$, the final state $| f \rangle$ will, in general, be a
different superposition of eigenstates from that of $| i \rangle$,
which was shown in~(\ref{eq:a4}). At high energies we may neglect
the real parts of the diffractive amplitudes. Then, for cross
sections at a given impact parameter $b$, we have
\bea \frac{d \sigma_{\rm tot}}{d^2 b} & = & 2 \:
{\rm Im} \langle i |T| i \rangle \; = \; 2 \: \sum_k
\: |a_{ik}|^2 \: F_k \; = \; 2 \langle F \rangle \label{eq:b1} \\
& & \nonumber\\
\frac{d \sigma_{\rm el}}{d^2 b} & = & \left | \langle i |T| i
\rangle \right |^2 \; = \; \left (
\sum_k \: |a_{ik}|^2 \: F_k \right )^2 \; = \; \langle F \rangle^2 \label{eq:b2}\\
& & \nonumber \\
\frac{d \sigma_{\rm el \: + \: SD}}{d^2 b} & = & \sum_k \: \left |
\langle \phi_k |T| i \rangle \right |^2 \; = \; \sum_k \:
|a_{ik}|^2 \: F_k^2 \; = \; \langle F^2 \rangle. \label{eq:b3} \eea
It follows that the cross section for the single diffractive
dissociation of a proton,
\be \label{eq:a7} \frac{d \sigma_{\rm SD}}{d^2 b} \; = \; \langle
F^2 \rangle \: - \: \langle F \rangle^2, \ee
is given by the statistical dispersion in the absorption
probabilities of the diffractive eigenstates. Here the average is
taken over the components $k$ of the incoming proton which
dissociates. If the averages are taken over the components of both
of the incoming particles, then (\ref{eq:a7}) is the sum of the
cross section for single and double dissociation.

Note that if all the components $\phi_k$ of the incoming
proton $| i \rangle$ were absorbed equally then the
diffracted superposition would be proportional to the incident one
and  the diffractive dissociation of the proton would be zero.  Thus if, at very
high energies, the amplitudes $F_k$ at small impact parameters are
equal to the black disk limit, $F_k = 1$, then diffractive
production will be equal to zero in this impact parameter domain
and so will only occur in the peripheral $b$ region. There are hints that such
behaviour is already starting to be approached in $pp$ (and $p\bar{p}$)
interactions at Tevatron energies. Hence the impact parameter
structure of diffractive dissociation and elastic scattering is drastically
different in the presence of strong $s$ channel unitarity effects.

At first sight, it appears that if we were to enlarge the number of eigenstates $|\phi_i\rangle$, then we could, in principle, include even high-mass dissociation, $pp \to X+p$, where one proton dissociates into a system $X$ of 
{\it high-mass} $M$. However, here we face the problem of double-counting when partons originating from dissociation of the beam and `target' initial protons overlap in rapidities. For this reason high-mass dissociation is usually described by ``enhanced'' multi-pomeron diagrams. We explain this procedure in Section \ref{sec:hm}.

\section{Structure of the pomeron   \label{sec:4}}

Originally the pomeron was introduced to describe the (approximately) constant behaviour of hadron-hadron cross sections with increasing energy in the (relatively) high energy domain. The pomeron was considered, formally, as just a pole in the complex angular momentum plane, see, for example, \cite{PeterCollinsbook}. Such an approach is close to the DL fit \cite{DL}.

Then it was recognized that we cannot restrict the description to only one pomeron exchange: $s$-channel unitarity generates multi-pomeron (non-enhanced) diagrams, while including $t$-channel unitarity we obtain more complicated `enhanced' multi-pomeron diagrams.

From the microscopic point of view the pomeron 
\footnote{see \cite{Levin} for a review.} was thought to be built from the sum of `ladder' type diagrams, originally from hadrons\footnote{The first multi-peripheral model \cite{AFS} sums up diagrams with $t$-channel pions which emit $\rho,~f_2,..$ mesons.}, but now, with the arrival of QCD, from gluons -- we speak of the BFKL or QCD pomeron.  The summation of such ladder diagrams never gives one individual pole, but a few poles (in the hadronic multi-peripheral case) or a series of poles, or even a cut, in the LO BFKL case \cite{book}. Therefore it looks plausible to consider a model in which we parametrise the vacuum exchange (effective pomeron) amplitude by a few poles.

This possibility was exploited in~\cite{L2} where
 the elastic total cross section was described by two pomerons; one with 
intercept 0.09, and a second with a much higher intercept $0.36$.  The residue of the latter (small size) pomeron has a relatively flat $t$-behaviour, so that it contributes mainly to the low $b$ region. Recall that accounting for the boundary condition in a low $k_t$ (confinement) region, the BFKL vacuum singularity becomes a series of the poles, which in the first approximation may be mimicked by two poles. The problem, however, is that in the case of BFKL the highest intercept pole corresponds to the `ground state' of the two $t$-channel gluon system and so should have the largest coupling to the proton, while to describe the data 
one needs this rightmost pole to have a relatively small coupling\footnote{However, it is not excluded that a large effective coupling of the lower intercept term is actually due to the summation over a large number of poles, each with relatively small intercepts and couplings.}.  That is at lower energies the major contribution is provided by the lower intercept pole.  So only above the Tevatron energy does the rightmost pole (with a large intercept) become noticeable.

More general analyses based on multi-channel eikonals for more than one pomeron,  
were performed e.g in KMR \cite{Ryskin:2009tj,KMR-s3,KMRLHC,Khoze:2013jsa},
GLM \cite{Maor,Gotsman:2013lya}, Ostapchenko \cite{Ost}, 
 Anisovich et al. \cite{anis,Anisovich:2013gxa}. 
Note that  ~\cite{Ost} includes two diffractive (G-W) eigenstates and two pomerons. Unfortunately these two pomerons were collected in a {\em single} pomeron propagator. So the natural possibility to have each pomeron with its own individual coupling  was not exploited.

The structure of the pomeron was exploited further in \cite{KMR-s3}, where the pomeron wave function was considered as a system of two $t$-channel gluons with the transverse momentum $k_t$.  The $k_t$ dependence of the couplings was included explicitly; that is we deal, not with just a few pomerons.  Instead we deal with the internal structure of the pomeron written in terms of its wave function which is a function of $k_t$. Numerically this is was mimicked by a set of pomeron states with different $k_t$, and we had the possibility to trace the BFKL-like diffusion as transitions between states with different $k_t$ during the evolution in rapidity.
We shall see in Section \ref{sec:global} that this $k_t$ dependence is crucial to understand the behaviour of the recent LHC elastic and diffractive data.

\subsection{Enhanced absorption: multi-pomeron diagrams}
Until now, we have 
accounted for the rescattering of the incoming partons only. However, from a microscopic point of view (both `hard' QCD  and  `soft')  pomeron exchange is described by a set of ladder-type diagrams. Thus we cannot exclude the rescattering of intermediate
partons (produced inside this ladder during the evolution). In terms of the RFT these effects are described by the triple- ($g_{3P}=g^1_2$) and the multi- ($g^n_m$) pomeron vertices, that is by the pomeron-pomeron interactions.  The vertex $g^n_m$ couples $m$ to $n$ pomerons.

\subsection{High-mass dissociation   \label{sec:hm}   }
\begin{figure}
\begin{center}
\includegraphics[height=6cm]{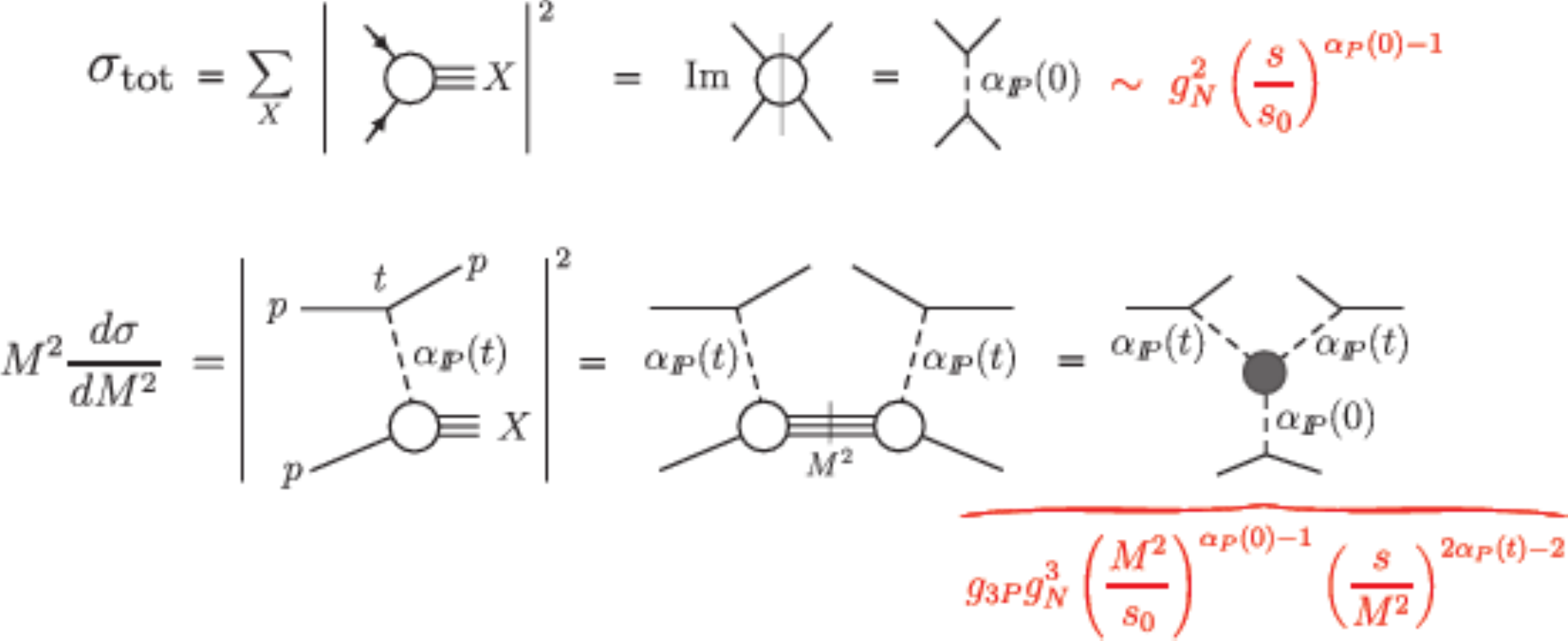}
\caption[*]{\sf Optical theorems for the total cross section and for high-mass diffractive dissociation. The expressions shown are for the {\it bare} amplitudes. At high energies these bare amplitudes have sizeable absorptive corrections, which are not shown in the figure.}
\label{fig:opt}
\end{center}
\end{figure}

\begin{figure} [h]
\begin{center}
\includegraphics[height=8cm]{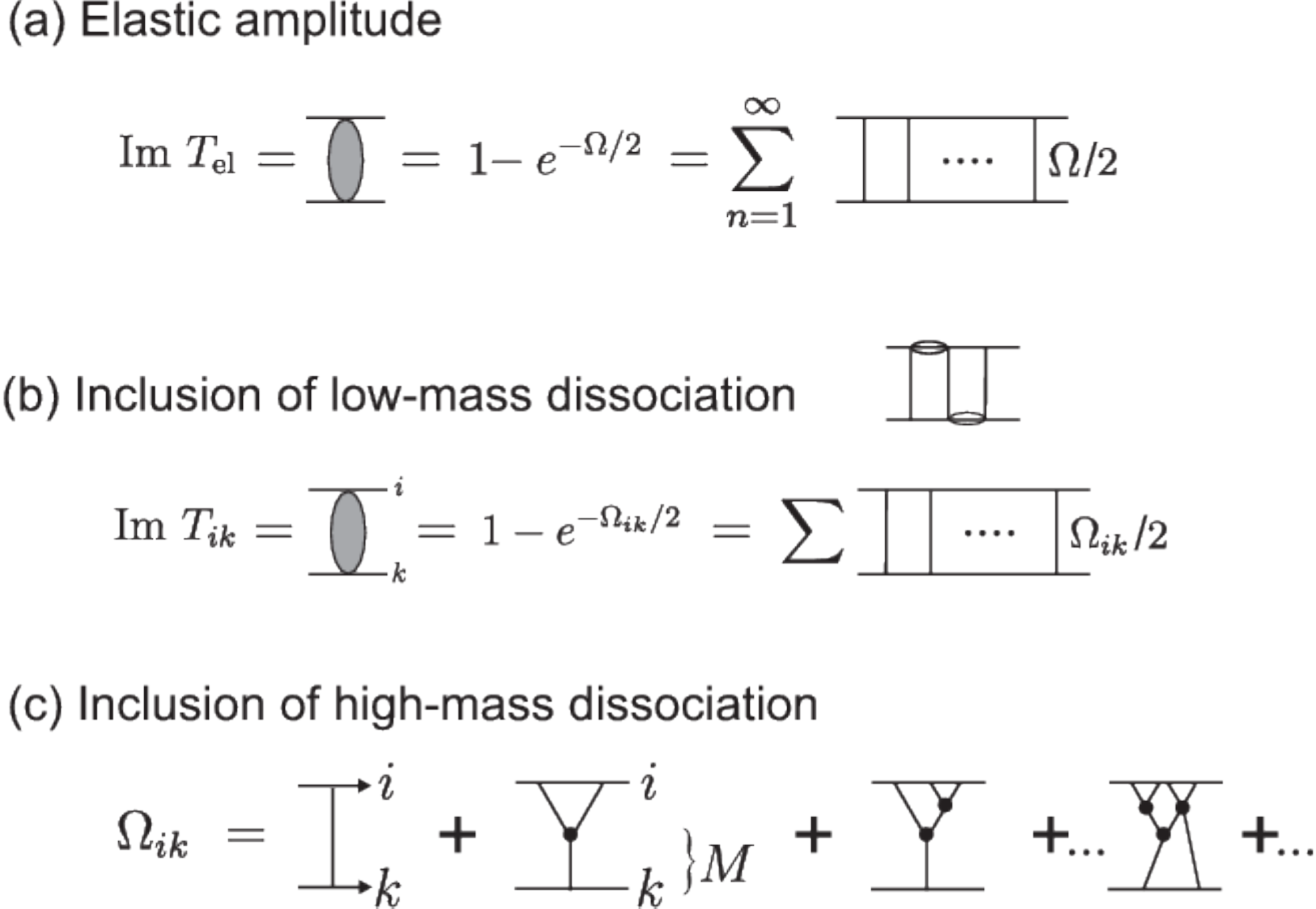}
\caption[*]{\sf (a) The single-channel eikonal description of elastic scattering; (b) the multichannel eikonal formula which allows for low-mass proton dissociations in terms of diffractive eigenstates $|\phi_i\rangle,~|\phi_k\rangle$; and (c) the inclusion of the multi-pomeron-pomeron diagrams which allow for high-mass dissociation. }
\label{fig:AA}
\end{center}
\end{figure}

 The first and simplest multi-pomeron diagram is the triple-pomeron graph, shown at the end of Fig.~\ref{fig:opt}, and again in Fig.~\ref{fig:AA}(c).  From Fig.~\ref{fig:opt}  we see, that by generalising the optical theorem to the dissociative process, $pp\to  p+X$, this triple-pomeron diagram applies when $X$ is a high mass system.
Indeed, in the absence of absorptive corrections, the
corresponding  $pp\to  p+X$ cross section is given by
\be
\frac{M^2 d\sigma_{\rm SD}}{dtdM^2}~=~\frac{g_{3P}(t)g_N(0)g_N^2(t)}{16\pi^2}~\left(\frac{s}{M^2}\right)^{2\alpha(t)-2}~\left(\frac{M^2}{s_0}\right)^{\alpha(0)-1},
\label{eq:3P}
\ee
where $g_N(t)$ is the coupling of the pomeron to the proton and $g_{3P}(t)$ is the triple-pomeron coupling.
The value of the coupling $g_{3P}$ is obtained from a triple-Regge analysis of lower energy data.  Mainly they are the data on proton dissociation taken at the CERN-ISR with energies from $23.5 \to 62.5$ GeV.

The problem, with the above determination of $g_{3P}$, is that the value obtained is actually an effective vertex with coupling
\be
g_{\rm eff}~=~g_{3P}~\langle S^2\rangle
\ee
which already includes the suppression $S^2(b)=\exp(-\Omega(b))$ -- the probability that no other secondaries, simultaneously produced in the same $pp$ interaction, populate the rapidity gap region denoted by the + sign in $pp \to p+X$, see (\ref{eq:S2}).  Recall that this survival factor $S^2$ depends on the energy of the collider.  Since the opacity $\Omega$ increases with energy, the number of multiple interactions, $N \propto \Omega$, grows\footnote{This is because at larger optical density $\Omega$ we have a larger probability of interactions.}, leading to a smaller $S^2$.  Thus, we have to expect that the naive triple-pomeron formula with the coupling \cite{abk,KKPT}, 
measured at relatively low collider energies will appreciably overestimate the cross section for high-mass dissociation at the LHC. A more precise analysis \cite{Luna} accounts for the survival effect $S^2_{\rm eik}$ caused by the eikonal rescattering of the fast `beam' and `target' partons.  In this way, a coupling $g_{3P}$ about a factor of 3 larger than $g_{\rm eff}$ is obtained, namely $g_{3P} \simeq 0.2g_N$, where $g_N$ is the coupling of the pomeron to the proton. The analysis of Ref. \cite{Luna} enables us to better allow for the energy dependence of $S^2_{\rm eik}$.  We will show how to include the absorptive corrections to formula (\ref{eq:3P}) when we come to describe the LHC data for high-mass diffractive dissociation.

Fig.~\ref{fig:AA} pictorially summarizes the extension of the description of the elastic amplitude, which enables  the related low- and high-mass proton dissociative processes to be included.  Notice the inclusion of other multi-pomeron graphs, besides the triple-pomeron graph, in diagram (c).

\section{The high energy diffractive data   \label{sec:data}}
We are now in a position to confront the `soft' data.  So let us summaise the present experimental situation.

At the moment, data for diffractive processes are available at 7 TeV. The most detailed data come from the TOTEM collaboration. TOTEM have measured the total and elastic cross sections (in a wide $t$ interval including the dip region)~\cite{TO1,TO}, the cross section of low-mass ($M_X <3.4$ GeV) diffractive
 single ($pp\to p+X$)~\cite{TO2} and double ($pp\to X_1+X_2$)~\cite{TO3} dissociation; and made preliminary measurements of high-mass single proton dissociation, $\sigma_{\rm SD}$, integrated over the three intervals of $M_X$: namely $(3.4,8);~(8,350);~(350,1100)$ GeV~\cite{TO4}. In addition we have the inelastic cross sections and the cross sections of events with a Large Rapidity Gap (LRG) measured by the ATLAS \cite{atl}, CMS \cite{CMSdiff} and ALICE \cite{ALICE} collaborations.  Moreover, we have valuable data on elastic and proton dissociation from experiments at the Tevatron \cite{EEE,cdfB,GM}.

\subsection{Potential puzzles and tensions in the data sets   \label{sec:items}}
Formally the diffractive dissociation data from different groups do not contradict each other, since they are measured for different conditions. However there appear to be several potential puzzles and tensions between the data sets. 
\begin{itemize}
\item First, it is not easy to accommodate simultaneously the TOTEM result for $\sigma_{\rm SD}$ and the yield of LRG events observed by  ATLAS/CMS. 

\item Moreover, the TOTEM $\sigma_{\rm SD}$ cross section looks too small in comparison with the value of $d\sigma_{\rm SD}/d\xi dt$ cross section measured by CDF at Tevatron energy, as given in \cite{GM}.
In particular, at $\sqrt s=1.8$ TeV, with a proton momentum fraction transferred through the pomeron of $\xi=1-x_L=0.01$  and $-t=0.05$ GeV$^2$, the CDF collaboration claim 
\be
d\sigma/d\ln\xi dt\simeq 2 ~{\rm mb/GeV^2},
\ee
 while TOTEM at $\sqrt s=7$ TeV  gives about\footnote{To obtain this estimate we have 
 divided the cross section (3.3 mb for single proton dissociation of {\em both} incoming protons) 
 measured in the central $8 <M_X < 350$ GeV interval by the size ($\Delta\ln 
 M^2_X=7.56$) of the rapidity interval, and accounted for the corresponding $t$-slope 
 ($B=8.5$ GeV$^{-2}$) observed by TOTEM~\cite{TO4}. Thus we obtain  $d\sigma_{\rm SD}/d\ln\xi 
 dt=(3.3~{\rm mb}/2/7.56)\times 8.5~$GeV$^{-2}\times\exp(-8.5\times 0.05)=1.2$ mb/GeV$^2$.} 
 1.2 mb/GeV$^2$, for the same mass of the diffractive state, $M_X\sim 100 - 200$ GeV. That is, TOTEM has a cross section about factor 1.7 smaller than CDF. On the other hand, naively, we would expect the value of the diffractive dissociation cross section to increase with energy.

\item Next the cross section $d\sigma_{\rm SD}/d\ln\xi$ in the first (3.4 to 8 GeV) $M_X$ interval is more than twice larger than that in the central interval. Of course, according to the triple-Regge formula, a pomeron intercept $\alpha_P(0)>1$ leads to an increase of the cross section when $\xi$ decreases, but by the same argument we have to observe a larger cross section at the LHC than at the Tevatron, for the same value of $M_X$, contrary to the data. 

\item An analogous problem is observed for low-mass dissociation, where the cross section, $\sigma_{\rm SD}^{{\rm low}M_X}$, was about 30\% of the elastic cross section at CERN-ISR and fixed target energies \cite{Kaid}, whereas it turns out to be only 10\% at the LHC~\cite{TO2}.
\item The `factorisation' relation between the observed elastic, single and double proton dissociation cross sections is intriguing, and appears not easy to explain. This puzzle merits discussion in a separate subsection, see Section \ref{sec:DDfac}.
\item To this list, we add the surprises in the elastic data mentioned in Section \ref{sec:surprise} 
\end{itemize}

We will seek an single explanation of all these puzzles by performing a global description to all the data.  The result is presented in Section \ref{sec:global}. But, first, let us explain the penultimate puzzle in more detail.

\subsection{Factorisation and Double Dissociation \label{sec:DDfac}}
 
 The recent TOTEM measurement of high energy double dissociation \cite{TO3} opens the way to study the relation between elastic, single dissociation and double dissociation cross sections.  Let us check this relation to see why it is a potential problem.
 
\begin{figure} 
\begin{center}
\vspace{-3.cm}
\includegraphics[height=10cm]{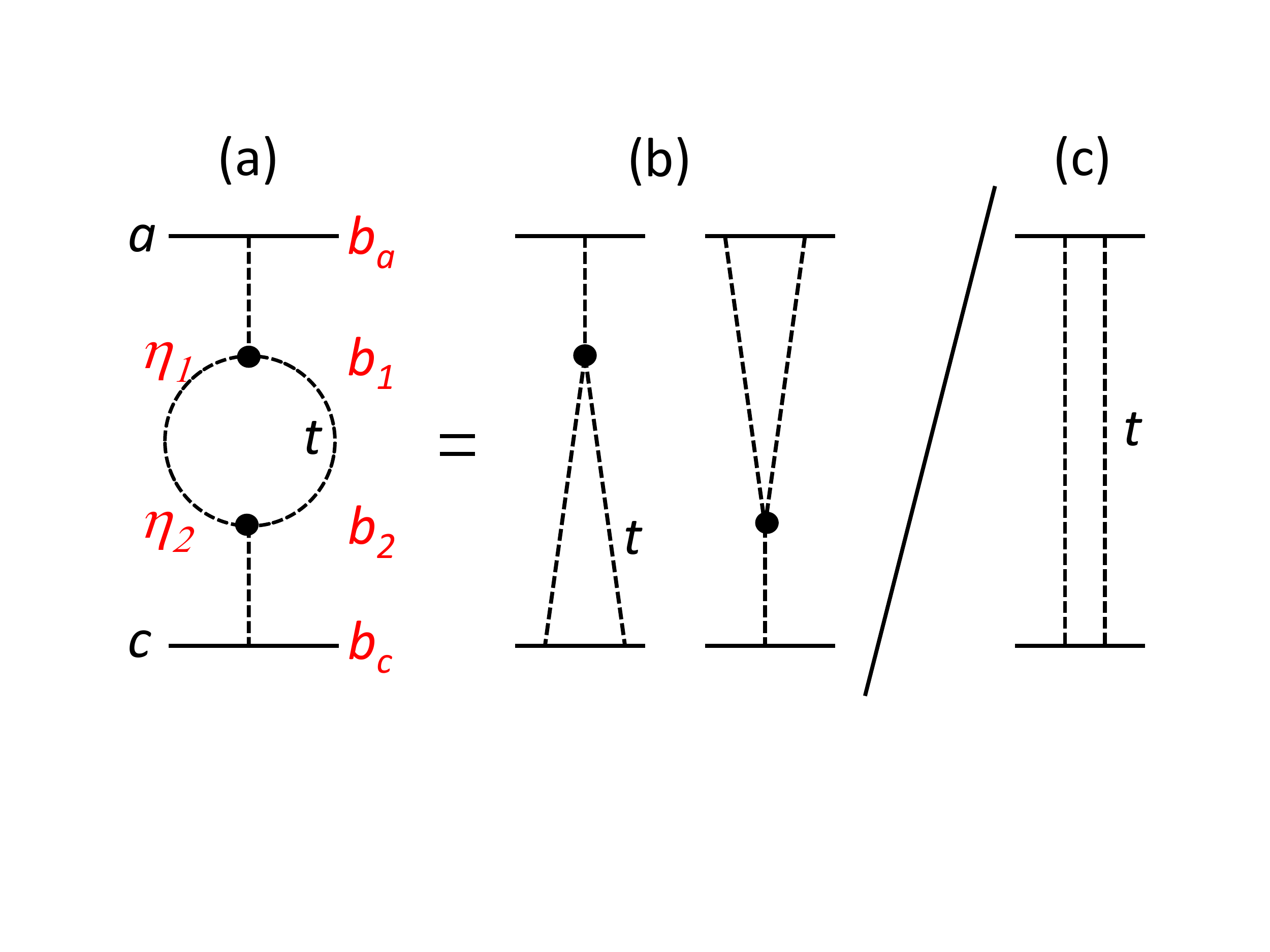}
\vspace*{-2.5cm}
\caption{\sf A pictorial representation of the naive factorization formulae of (\ref{eq:1}) and (\ref{eq:2}), resulting from the simplest pomeron exchange diagrams for (a) DD, (b) SD*SD and (c) elastic $ac$ scattering. It is convenient to evaluate the dissociation cross sections in impact parameter space, so we also show the variables $b_i$. }
\label{fig:3}
\end{center}
\end{figure}

 Within the framework of RFT, the simplest Reggeon diagram which describes the cross 
 section of high-mass diffractive double dissociation at high energies is the pomeron exchange diagram shown in 
Fig. \ref{fig:3}(a). As clear from Fig. \ref{fig:3}, it is natural to expect the 
 factorization relation
\be
\frac{d\DD}{dtd\eta_1d\eta_2}~~=~~\frac{d\SD}{dtd\eta_1}~\frac{d\SD}{dtd\eta_2}~/~\frac{d\el}{dt}.
\label{eq:2}
\ee
to be valid. Note that relation (\ref{eq:2}) is written for the differential cross section for some fixed value of the square of the momentum transfer $t$, and not for the cross sections integrated over $t$. The corresponding naive integrated factorisation relation is
\be
\DD~=~\frac{(\SD)^2}{\el}~~~~~~~~~~{\rm or}~~~~~~~~~~\frac{\DD~\el}{(\SD)^2}~=~1,
\label{eq:1}
\ee
where here $\SD$ is the single dissociation cross section from {\it one} proton, not the sum of both dissociations.
Before we compare the factorisation relation with the cross sections obtained by the  TOTEM collaboration  at $\sqrt{s}=7$ TeV, we must include some obvious violations expected for the naive form (\ref{eq:1}).

First, the relation is violated by the different $t$-slopes, $B$, of the elastic, single and double dissociation cross sections. Indeed, at 7 TeV the corresponding slopes are:
 $B_{\rm el}\simeq 20$ GeV$^{-2}$\cite{TOTEM2,TO}, $B_{\rm SD}\simeq 10$ GeV$^{-2}$ for the lowest mass interval\footnote{To be specific, the preliminary values of the slopes observed by TOTEM \cite{TO4} in their three mass intervals are $B_{\rm SD}= 10.1,~8.5,~6.8~$ GeV$^{-2}$ respectively, with 15\% errors.} in \cite{TO4}, and the 
 estimated slope 
 \be
 B_{\rm DD}~\simeq ~4B_{3P}+2\alpha'_P|\eta_1-\eta_2|~=~ 3.8 
 ~\GeV^{-2}.
 \ee
 corresponding to the TOTEM experimental kinematics with 
 $|\eta_1-\eta_2|\sim 10$. For the estimate of $B_{\rm DD}$ we take the   
value  $\alpha'_P=0.05$ GeV$^{-2}$
obtained in Section \ref{sec:pom} to describe the elastic proton-proton cross section, 
 and we put the slope of the triple-pomeron vertex $ B_{3P}=0.7$ GeV$^{-2}$ (which is consistent with the H1 data on elastic and proton dissociative $J/\psi$ photoproduction at HERA \cite{H1}). Thus we already expect a violation of the naive relation (\ref{eq:1}) by a factor
 $B_{\rm SD}^2/B_{\rm el}B_{\rm DD}\sim 1.36$.

 More serious are the role of the 
 eikonal rapidity gap survival factors $S^2_{\rm eik}$. Both the single and the double 
 dissociative cross sections are suppressed by $S^2$. However, $S^2_{\rm SD}$ enters 
 (\ref{eq:1}) as the square of $S^2_{\rm SD}$, while $S^2_{\rm DD}$ enters as the first power. 
The elastic 
scattering  cross section, which results from unitarity, has no explicit $S^2$ suppression, but, after accounting for the multi-pomeron diagrams, its value becomes less than that given by  single pomeron pole exchange, that is 2Im$T(b)<\Omega (b)$, see (\ref{eq:1i}). Using the `elastic' parameters obtained in the global fit to the `soft' data (given in Section \ref{sec:elas} below) we find a suppression of $d\sigma/dt|_{t=0}$ by a factor of about $ 6.8$. Moreover,   double dissociation occurs typically at somewhat larger values of the impact parameter, $b$, so $S^2_{\rm DD}>S^2_{\rm SD}$, see, for example, \cite{KMR18,Kaidalov:2003gy}.  These observations all lead to the left-hand-side being larger than the right-hand-side of (\ref{eq:1}). 

Thus, it is not a surprise to find sizeable breaking of naive factorisation. The question is whether we can account for the actual observed size of the breaking. Using the global model, described below, we find $S^2_{\rm SD}\simeq 0.08$ and a twice larger $S^2_{\rm DD}\simeq 0.16$. Thus, including the suppression of the elastic cross section and the slope factor, our estimate so far is
\be
\frac{\DD~\el}{(\SD)^2}~\simeq~\frac {1.36}{6.8} ~\frac{0.16}{(0.08)^2}~\simeq 5.0.
\label{eq:rth}
\ee
 On the other hand, the TOTEM data give a much smaller violation of factorisation
 \be
\frac{\DD~\el}{(\SD)^2}~\simeq~\frac{0.116 \times 25}{(0.9)^2}~\simeq 3.6,
\label{eq:rexp}
\ee
 where here we use $\DD=0.116$ mb \cite{TO3},
 $\SD=1.8/2=0.9$ mb \footnote{The TOTEM result of 1.8 mb corresponds to single dissociation of {\em both} 
 protons, in the {\it same} rapidity interval as used for their measurement of $\DD$.}\cite{TO4} and $\el=25$ mb\cite{TOTEM2,TO}.  We will have to see if our global description will also resolve this difference.

\section{Global description of `soft' high-energy data \label{sec:global}}
The puzzles in the energy behaviour of the diffractive cross sections, that we discussed above, result from attempting to describe the data within the eikonal framework based on pomeron exchange, together with multi-pomeron (absorptive) corrections.  These suggest that there is a missing, physically-motivated, ingredient in the model.  What can it be?

We find that the missing ingredient, such that all these puzzles may be explained semi-quantitatively, is that the values of the pomeron couplings should not be fixed, but allowed to decrease with energy due to the growth of $k_t$ of the intermediate partons along the pomeron exchange ladder\footnote{Note that, in conventional RFT, it was assumed that all the transverse momenta are
limited (and small). On the other hand, the mean transverse momentum of secondaries
 increases with energy (due to the stronger absorption of the low 
$k_t$ partons). Thus there is a physics reason to revise this assumption and to include some rapidity/energy dependence of $k_t$ into the RFT approach.}.
That is, we study the role of the energy (and rapidity) dependence of the transverse momenta of gluons which form the pomeron and find that  this dependence enables us to reproduce the new qualitative properties observed at the LHC.

Recall that the mean $p_t$ of secondaries, $\langle p_t \rangle$, indeed grows with energy. In particular, the tuning of the Pythia8 Monte Carlo \cite{P8} required the introduction of an energy dependent infrared cutoff $k^2_{\rm min} \propto s^{0.24}$.

Bearing in mind the relatively small value of the triple-pomeron and the multi-pomeron vertices, we start with the simplest Reggeon diagrams, then include the absorptive (gap survival) effects caused by the eikonal and consider the role of the increasing transverse momenta which lead to the decrease of the
pomeron (and the multi-pomeron) couplings ($\propto 1/k_t$) with increasing $k_t$. To make the discussion more transparent we will not include explicitly the enhanced diagrams (which account for the rescattering of the intermediate ladder partons). The role of these
diagrams is mainly to renormalize (decrease) the intercept of the original (bare) pomeron and to enlarge the characteristic transverse momenta, due to the stronger absorption of the partons with low $k_t$. Therefore, below, we will use renormalized  parameters of the pomeron trajectory\footnote{The major role of the enhanced diagrams is the renormalisation of the pomeron trajectory and vertices. The remaining effects of the multi-pomeron interactions (that is, the multi-pomeron vertices $g^m_n$) is not so strong. So at present we will include the minimal number of such vertices, $g^m_n$, needed to obtain high-mass single or double dissociation.},
 obtained by fitting to the data,
together with a reasonable assumption about the energy and rapidity behaviour
of $k_t$.

 In the following subsections we describe the fit to the various types of diffractive data. Although these discussions may seem to be self-contained analyses, we emphasize that they are just parts of a simultaneous `global' description of all types of high-energy diffractive data.  We give a discussion in Section \ref{sec:6}, together with a summary of model predictions of high energy diffractive observables.

\subsection{Description of elastic scattering \label{sec:elas}}
In terms of the G-W framework, the differential elastic cross section takes the form
\be
\frac{d\sigma_{\rm el}}{dt}~=~\frac{1}{4\pi}  \left| \int d^2b~e^{i\q_t \cdot \b} \sum_{i,k}|a_i|^2 |a_k|^2~(1-e^{-\Omega_{ik}(b)/2}) \right|^2,
\label{eq:elass}
\ee
where $-t=q_t^2$, and in which both of the incoming proton states are expressed as a linear sum over the diffractive eigenstates, $|p\rangle = \sum_i a_i |\phi_i\rangle$.  The opacity is driven by one-pomeron-exchange (between states $\phi_i$ and $\phi_k$ in the $b$-representation)
\be
\label{eq:ob}
\Omega_{ik}(s,b)=\int\frac{d^2q_t}{4\pi^2}\Omega_{ik}(s,q_t)e^{i\q_t \cdot \b}
\ee
with
\be
\Omega_{ik}(s,q_t)=g_i(t) g_k(t) \left(\frac s{s_0}\right)^{\alpha_P(t)-1},
\label{eq:ot}
\ee
where $g_i$ is the coupling of the $\phi_i$ eigenstate to the pomeron.
We use a two-channel eikonal; that is, two G-W diffractive eigenstates $i,k=1,2$. The normalization, Im$T=s\sigma$, is such that the pomeron-nucleon couplings 
\be
g_i=\gamma_i\sqrt{\sigma_0}F_i(t),
\label{eq:gamma}
\ee
where the form factors satisfy $F_i(0)=1$. Thus the cross section for the interaction of eigenstates $\phi_i$ and $\phi_k$, via one-pomeron-exchange, is
\be
\sigma_{ik}=\sigma_0 \gamma_i \gamma_k (s/s_0)^\Delta.
\ee
The form factors are parametrized as
\be
F_i(t)={\rm exp}((-b_i(c_i-t))^{d_i}+(b_ic_i)^{d_i}).
\label{eq:ff}
\ee
The $c_i$ term is added to avoid the singularity $t^{d_i}$ in the physical region of $t<4m^2_{\pi}$. Note that $F_i(0)=1$. 

The parameters $b_i,~c_i,~d_i$, together with the intercept and slope of the pomeron trajectory are tuned to describe the elastic scattering data, paying particular attention to the energy behaviour of 
low mass dissociation cross section. 
We first discuss the description of the elastic data. 

In order to correctly describe the dip region we must include the real part of the amplitude. We use a dispersion relation. For the even-signature pomeron-exchange amplitude this means
\be
A~\propto ~s^{\alpha} + (-s)^{\alpha} ~~~~~~~{\rm and~so~we ~have} ~~~~~~~\frac{{\rm Re}~A}{{\rm Im}~A}={\rm tan}(\pi(\alpha -1)/2).
\label{eq:RE}
\ee 
That is, we restore the complex $\Omega (b)$ by means of the usual signature factor
\be
\eta_P (t)~=~i-{\rm cotan}(\pi\alpha_P/2)~=~i+{\rm tan}(\pi (\alpha_P-1)/2),
\ee
which follows from (\ref{eq:signature}). Formula (\ref{eq:RE}) is transformed into $b$-space, so that the complex opacities, $\Omega_{ik}(b)$ in (\ref{eq:ob}) can be constructed. For each value of $b$, that is for each partial wave $l$, we calculate $\alpha$ and determine Re$~A$ from (\ref{eq:RE}).  

\begin{figure} 
\begin{center}
\vspace*{-10.0cm}
\includegraphics[height=21cm]{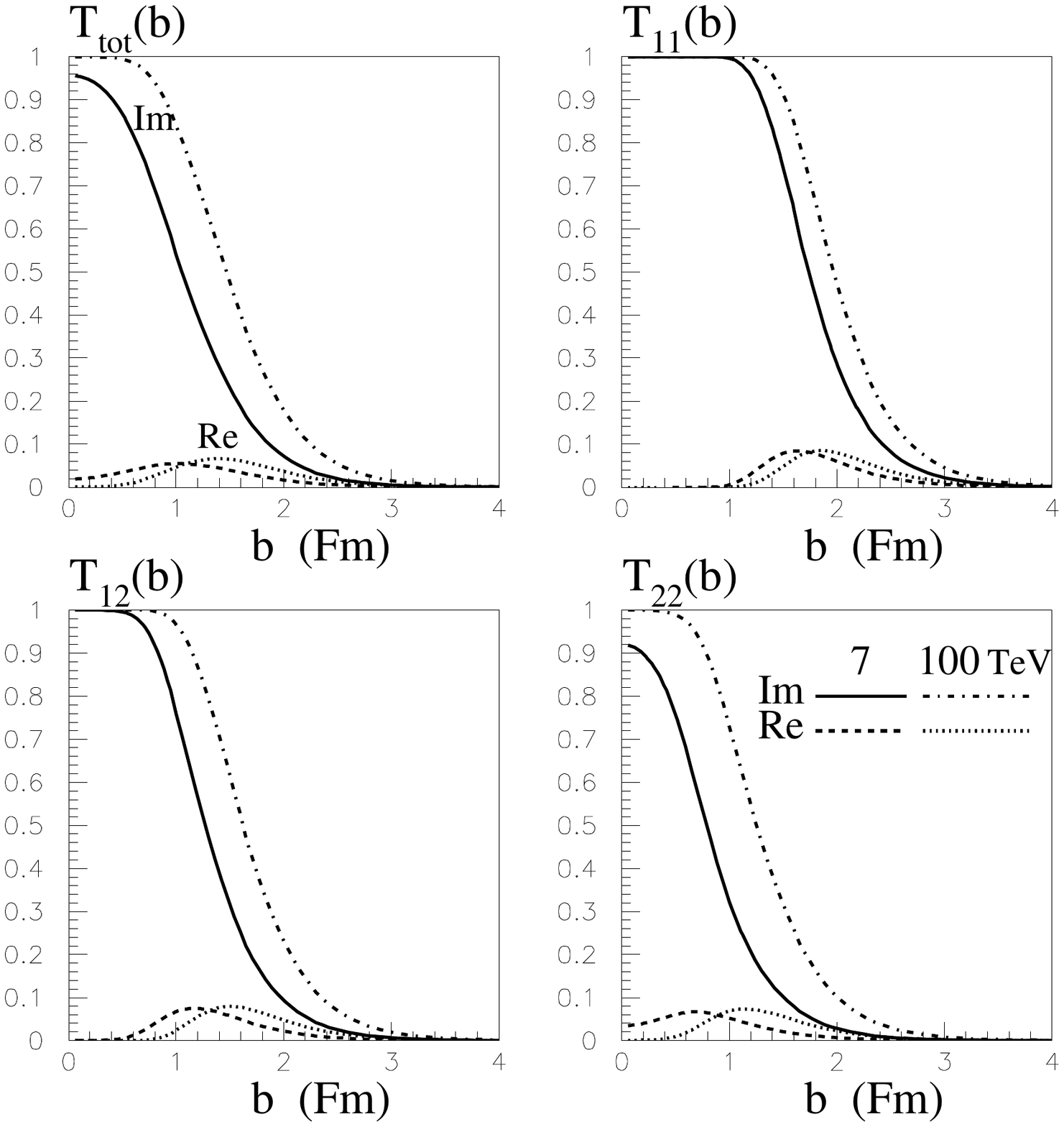}
\vspace*{-0.5cm}
\caption{\sf The real and imaginary parts of the amplitude of elastic proton-proton scattering, $T_{\rm tot}(b)$, together with the contributions arising from the individual diffractive eigenstates $\phi_i$.  That is $T_{\rm tot}=|a_1|^2 |a_1|^2T_{11}+2|a_1|^2 |a_2|^2T_{12} +|a_2|^2 |a_2|^2T_{22}  $, as follows from the summation over $i,k$ in (\ref{eq:elass}). }
\label{fig:t-b}
\end{center}
\end{figure}

The resulting real and imaginary amplitudes, $T(b)$, are shown in Fig. \ref{fig:t-b} for elastic proton-proton scattering  and for the individual eigenstates $\phi_i$ contributions at $\sqrt{s}=7$ and 100 TeV.
Note that for the whole elastic amplitude, $T_{\rm tot} (b)$, we still do not reach the black disk limit at 7 TeV. Even at 100 TeV, it is reached only for small $b$, namely $b<0.4$ fm. On the other hand, the large size (and large cross section) eigen-amplitude $T_{11}$ already reaches the black disk limit at rather large $b$, namely at $b\lapproxeq 1$ fm for $\sqrt{s}=$ 7 TeV. The real parts of the amplitudes are small at large $b$ and as $b \to 0$; their maximum corresponds to the interval where Im$T$ is most steep.

In order to reproduce the cross section in the diffractive dip region  we find that the form factors, (\ref{eq:ff}) have to have powers $d_1=0.52$ and $d_2=0.51$, close to the form used long ago by Orear {\it et al.}, $F={\rm exp}(-b\sqrt{t})$~\cite{Or}. The values of the other parameters are
$c_1=0.35,\, c_2=0.25,\, b_1=4.7,\, b_2=4.1$ in GeV units.
In addition we take $|a_1|^2=0.265$, with $|a_2|^2=1-|a_1|^2$, and $\sigma_0\equiv (g_N(0))^2=57$ mb, where $g_N(t)$ is the proton-pomeron coupling\footnote{The explicit values of the factors $\gamma_i$, and the parametrisation of their energy behaviour, will be described in the next subsection, since the dispersion of the $g^N_i$ couplings (caused mainly by the $\gamma_i$) is the major factor controlling the probability of low-mass dissociation.}.
  
\begin{figure} 
\begin{center}
\vspace*{-6.0cm}
\includegraphics[height=21cm]{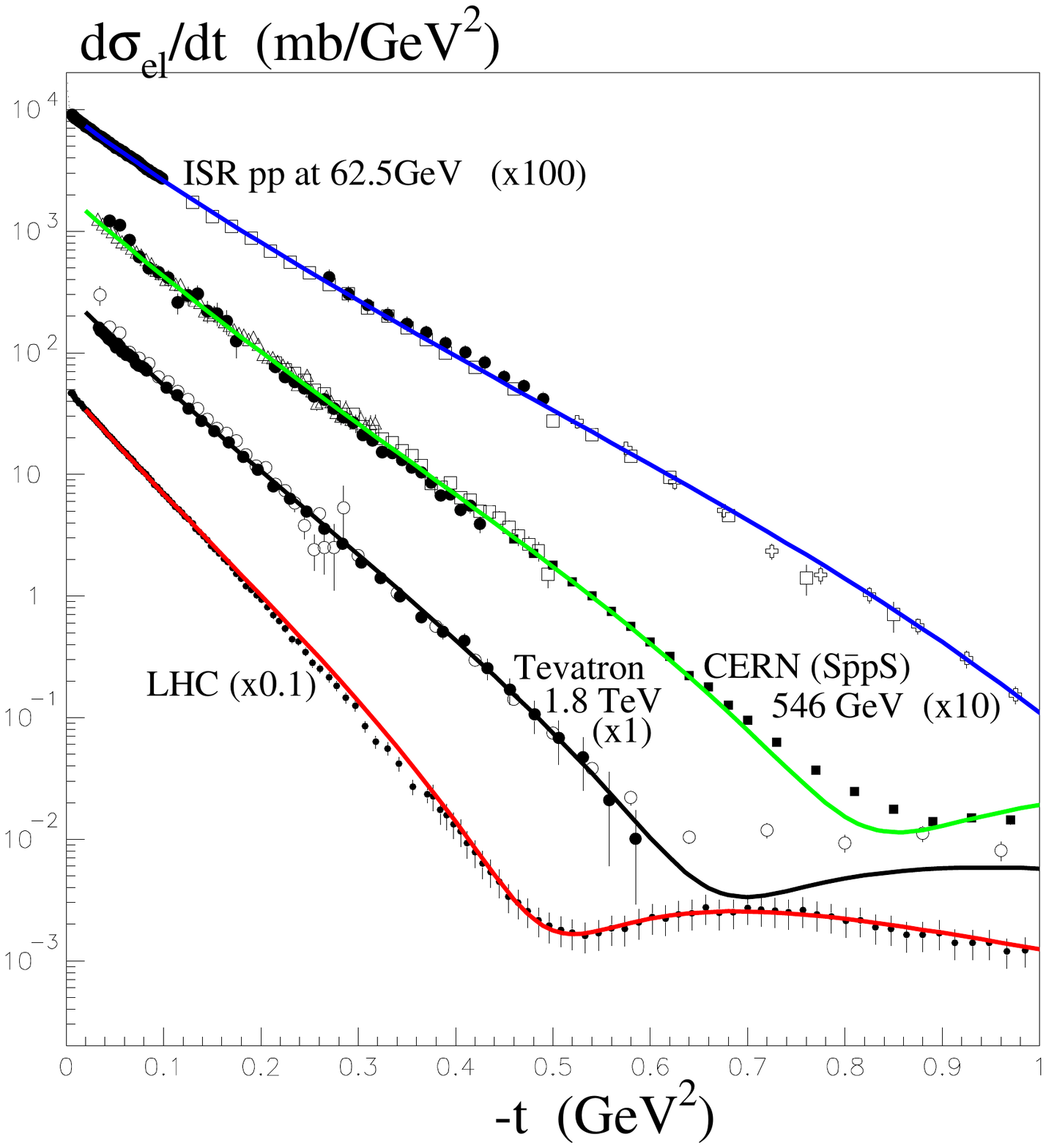}
\vspace*{-0.5cm}
\caption{\sf The description of $pp$ or ($p{\bar p}$) elastic data. The references to the pre-LHC elastic data can be found in \cite{SppS,Augier:1993sz,Arnison:1983mm,cdfB,EEE,Kwak:1975yq,
Amaldi:1976yf,Baksay:1978sg,Amaldi:1979kd,Bozzo:1985th,Abazov:2012qb}.
Here LHC refers to 7 TeV and the data are from \cite{TO1,TO}.
}
\label{fig:A}
\end{center}
\end{figure}

  The resulting description of the elastic  data is shown in Fig. \ref{fig:A}.
  The description of the proton-antiproton scattering at large $|t|>0.6$ GeV$^2$  is not perfect. This may be caused by the fact that we do not include secondary reggeon contributions. We also are not considering here a possible Odderon exchange contribution. 
 
\subsection{Description of low-mass dissociation}
The next part of the `global' description that we discuss is low-mass dissociation.  Here the experimental information is a puzzle in that the cross section
$\sigma_{\rm D}^{{\rm low}M_X}$ goes from about $2-3$ mb at the CERN-ISR energy\footnote{The 
relevant experimental references are \cite{Baksay:1974mn,Webb:1974nb,
Baksay:1976hb,deKerret:1976ze,Mantovani:1976vz}}
of 62.5 GeV to only $2.6\pm 2.2$ mb at 7 TeV at the LHC \cite{TO2}. Thus $\sigma_{\rm D}^{{\rm low}M_X}$ is about 30$\%$ of $\sigma_{\rm el}$ at 62.5 GeV and only about 10$\%$ at 7 TeV, whereas we would expect these percentages to be about the same for single pomeron exchange.
This problem was discussed in \cite{KMRLHC}, and its resolution involves more understanding of the decomposition of the G-W diffractive eigenstates $|\phi_i\rangle$.

The Good-Walker framework was discussed in Section \ref{sec:GW} for single proton dissociation. The generalisation to double dissociation is straightforward. For completeness we give the full expressions for the elastic and the `total' low-mass diffractive cross sections (analogous to (\ref{eq:b2}) and (\ref{eq:b3}) respectively)
\bea
\sigma_{\rm el}~=~  \int d^2b \left|~\sum_{i,k}|a_i|^2 |a_k|^2~(1-e^{-\Omega_{ik}(b)/2}) \right|^2,\\
\sigma_{\rm el+SD+DD}~=~  \int d^2b ~\sum_{i,k}|a_i|^2 |a_k|^2 ~\left|(1-e^{-\Omega_{ik}(b)/2}) \right|^2,
\eea
where SD includes the single dissociation of both protons.
So the low-mass diffractive dissociation cross section is
\be
\sigma_{\rm D}^{{\rm low}M}~=~\sigma_{\rm el+SD+DD}-\sigma_{\rm el}.
\ee

We are now ready to resolve the puzzle of the energy dependence of $\sigma_{\rm D}^{{\rm low}M_X}$. The pomeron-$|\phi_i \rangle$ coupling, $g_i$, 
 is driven by the impact parameter separation, $\langle r\rangle$, between the partons in the $|\phi_i\rangle$ state. The well known example is so-called colour transparency, where the cross section $\sigma\propto \alpha_s^2\langle r^2\rangle$~\cite{CT1,CT2,CT3,CT4}. However, if the transverse size of the pomeron, ($\sim 1/k_t$),  becomes much smaller than this separation, then the cross section (and coupling) will be controlled by the pomeron size, that is by the characteristic $k_t$ in the pomeron ladder. In this limit $\sigma\propto 1/k^2_t$. Such behaviour is clearly seen within the Low-Nussinov model \cite{Low,Nus}, where the pomeron is represented by the two-gluon exchange amplitude, see Appendix B. Therefore it is natural to choose the following parametrization for the pomeron-$|\phi_i\rangle$ couplings
\be
\gamma_i\propto \frac 1{k^2_P+k^2_i}, 
\label{eq:gam-i}
\ee 
where the $\gamma_i$ are defined in (\ref{eq:gamma}), with the normalization $(\gamma_1+\gamma_2)/2=1$. Here $k_P$ is the characteristic transverse momentum inside the pomeron, which we expect to behave as
\be 
k^2_P= k^2_{P0}\left(\frac{s x^2_0}{s_0}\right)^D\ .
\label{eq:D}
\ee 
In other words, during the evolution in $\ln(1/x)$, due to the BFKL 
diffusion in $\ln k^2_t$ \cite{Lipatov}, the square of the characteristic momentum $k^2_P$ grows approximately as a power $D$ of $1/x$. Of course, we do not expect that the whole available $\ln(1/x)$ (rapidity) space will be subject to  diffusion. Rather, we assume, that as $x$ decreases, the diffusion starts from some relatively low $x=x_0$ parton with $x_0=0.1$. That is, the rapidity space available for the $\ln k^2_t$ diffusion is not  $\ln(s/s_0)$, but is diminished by $\ln(1/x_0)$ from both sides. (As usual we use  $s_0=1$ GeV$^2$.) The typical transverse momentum of this (starting) parton, inside the state $\phi_i$, is denoted by $k_i$ in (\ref{eq:gam-i}). In our `global' model description we take $D=0.28$. The value of $D$ is related to the $s^\Delta$ behaviour, with $\Delta=0.2-0.3$, of resummed BFKL, which is mentioned in Section \ref{sec:pom} below. However, the relation is not direct. Rather, it is some approximation of the resummed BFKL diffusion in $\ln k_t$. For this reason we keep $D$ as a free parameter.

The parametrisation of $\gamma_i$ in (\ref{eq:gam-i}) is such that at very large energies all the $\gamma_i$ tend to the same value, so the dispersion shown in (\ref{eq:a7}) decreases leading to a smaller probability of low-mass proton dissociation, while at lower energies we tend to the naive expectation $\gamma_i \propto 1/k^2_i$. Actually the value of the additional transverse momenta $k_P$ in (\ref{eq:gam-i}) turns out to be rather small in the fit to the data --  $k_P/k_1=0.35$ and $k_P/k_2=0.17 $ at $\sqrt s=1800$ GeV.  Nevertheless the dissociation is slowed sufficiently with increasing energy such that we achieve values of the cross section $\sigma_{\rm D}^{{\rm low}M_X}$ which are compatible with the data -- namely, we find the model gives 2.6 mb at $\sqrt{s}=62.5$ GeV, and 3.8 mb at $\sqrt{s}=7$ TeV.  The latter value is consistent with the TOTEM measurement of 2.6$\pm$2.2 mb.

\begin{figure} 
\begin{center}
\vspace*{-7.5cm}
\includegraphics[height=20cm]{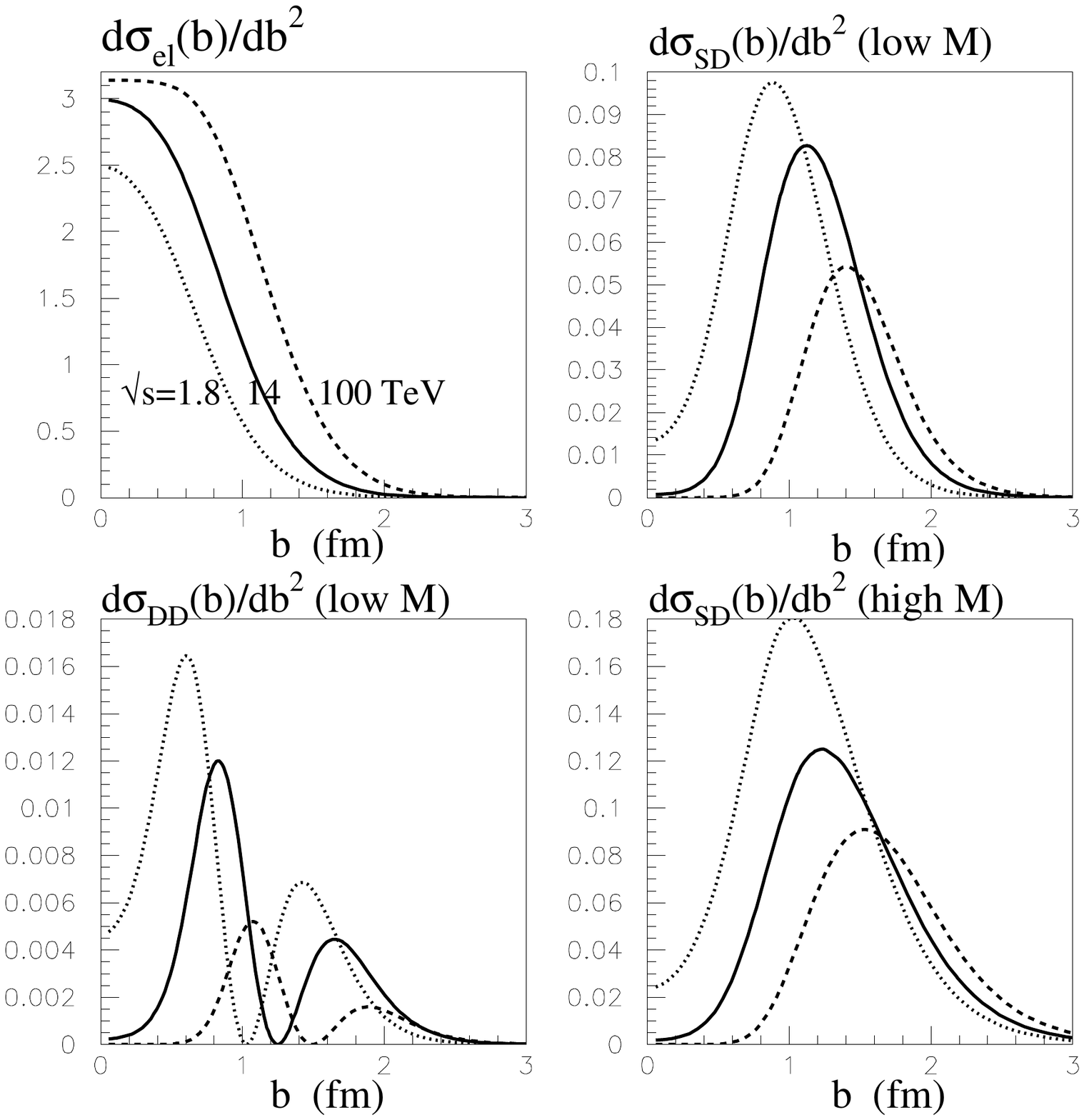}
\vspace*{-.5cm}
\caption{\sf The cross sections for elastic and diffractive dissociation as functions of the impact parameter, $b$, for $\sqrt{s}=1.8,~14$ and 100 TeV. The cross for high-mass dissociation is integrated over the same central rapidity interval as that selected by TOTEM, which corresponds to $M_{\rm diss}=(8,350)$ GeV at $\sqrt{s}=7$ TeV. $\SD$ is calculated for the dissociation of {\it one} proton. }
\label{fig:sig-b}
\end{center}
\end{figure}

The impact parameter distributions of the elastic and proton dissociation cross sections are shown in Fig. \ref{fig:sig-b}. Recall that each value of $b$ corresponds to a definite incoming partial wave $l=b\sqrt{s}/2$. Unlike the elastic case, the probability of dissociation vanishes as $b\to 0$ at high energies when the elastic amplitude becomes close to the black disk limit. Therefore dissociation comes mainly from the edge of the disk. The $b$ dependence of the cross section for dissociation of both protons, $\DD$, has a more complicated structure due to the interference between the different G-W eigenstates. 

The observation that the $\gamma_i\to 1$ at high energy, also means that the growth of the total cross section speeds up in the CERN-ISR $\to$ LHC $\to$ 100 TeV interval. In particular, if we fix the values of $\gamma_i$ to those at the CERN-ISR energy (say, to the values of $\gamma_i$ at $\sqrt s=50$ GeV) we obtain for $\sqrt s=1.8,\ 7$ and 100 TeV the values of $\sigma_{\rm tot}=74,\ 91$ and 134 mb respectively (instead of the values 77, 99 and 166 mb of the original version presented above). These numbers demonstrate an important new fact. That is, that the energy dependence of the total and elastic cross sections is not only driven by the parameters of the pomeron trajectory, but also by the energy behaviour of the factors $\gamma_i$; in other words by the decomposition of the proton-pomeron coupling between the different Good-Walker eigenstates. Note that this `acceleration' of the total cross section growth, due to the variation $\gamma_i \to 1$, takes place in only one energy interval. We are fortunate to observe it in just the  $Sp\bar{p}S$ -- LHC collider interval.

\subsection{Parameters of the `effective' pomeron trajectory \label{sec:pom}}

As mentioned, for simplicity, in the present approach we do not account explicitly for enhanced absorptive effects, which would renormalize the pomeron trajectory. Instead, we deal with an effective renormalized pomeron. Therefore  it is not surprising that the value $\Delta=0.12$ found for the effective pomeron is {\it larger} than 0.08 (the value obtained when the amplitude was parametrized by one-pole-exchange without any multi-pomeron corrections
 \cite{DL}), but is {\it smaller} than the intercept, $\Delta\sim 0.2\ -\  0.3$,
 expected for the bare pomeron of the resummed NLL$(1/x)$ BFKL approach 
 \cite{resum1,resum2,resum3}.  Indeed,
in comparison with the simple model, we explicitly account for the non-enhanced eikonal absorption which suppresses the growth of the amplitude with energy.  Therefore to describe the same data we need a larger intercept ($\Delta=0.12$).  On the other hand, since we do not explicitly include the enhanced diagrams (which would also slow down the growth of the cross section in the eikonal approach)  we anticipate a smaller effective intercept than that given by resummed BFKL.  Similar arguments apply to the slope of the effective trajectory, leading to a value\footnote{Besides the constant slope, $\alpha'$, of the pomeron trajectory, we insert the $\pi$-loop contribution as proposed in \cite{AG}, and as implemented in \cite{KMR18}} ($\alpha'=0.05 ~\GeV^{-2}$)  intermediate between the BFKL prediction ($\alpha' \gapproxeq 0$) and the old one-pole parametrization \cite{DL} ($\alpha'=0.25~ \GeV^{-2}$).

\subsection{High-mass single proton dissociation}
The framework for calculating high-mass dissociation was presented in Section \ref{sec:hm}.  The cross section was given in (\ref{eq:3P}), in the absence of absorptive corrections.
\begin{figure} 
\begin{center}
\vspace*{-3.5cm}
\includegraphics[height=9cm]{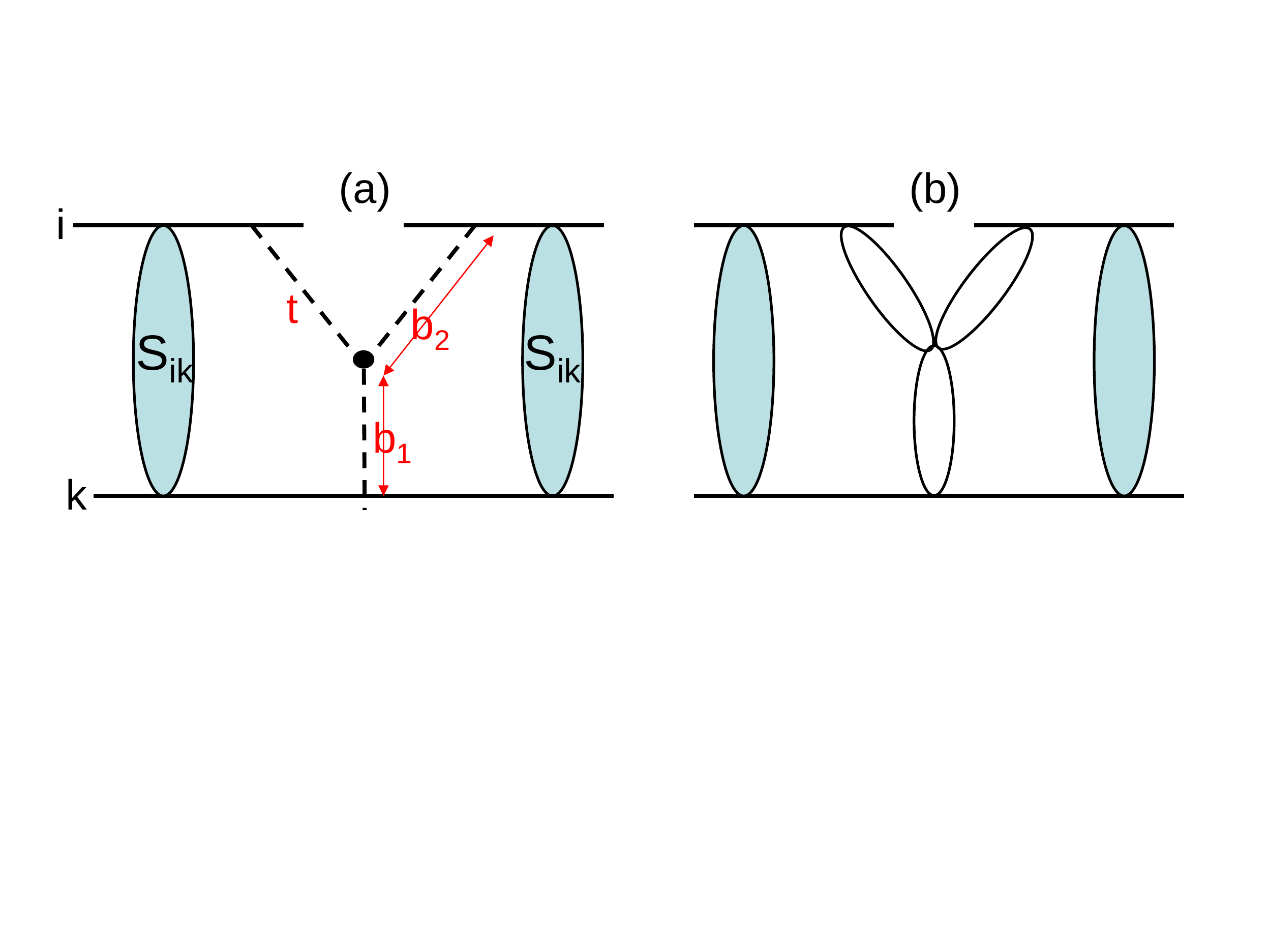}
\vspace*{-4.5cm}
\caption{\sf (a) A schematic diagram showing the notation of the impact parameters arising in the calculation of the screening corrections to the triple-pomeron contributions to the cross section; (b) a symbolic diagram of multi-pomeron effects. }
\label{fig:3Rb}
\end{center}
\end{figure}

To account for the absorptive effect, it is easier to work in the impact parameter, $b$, representation. 
To do this we follow the procedure of Ref. \cite{Luna}. We first take Fourier transforms with respect to the impact parameters specified in Fig. \ref{fig:3Rb}(a). Then (\ref{eq:3P}) becomes 
\be
\frac{M^2 d\sigma_{ik}}{dtdM^2}~=~A\int\frac{d^2b_2}{2\pi}e^{i\vec{q}_t \cdot \vec{b}_2} \Omega_i(b_2)\int\frac{d^2b'_2}{2\pi}e^{-i\vec{q}_t \cdot \vec{b}'_2} \Omega_i(b'_2 )\int\frac{d^2b_1}{2\pi} \Omega_k(b_1),
\label{eq:3Rb}
\ee
where $\Omega_i(b)$ is 
the opacity corresponding to the interaction of eigenstate $\phi_i$ with a intermediate parton placed at the position of the triple-pomeron vertex, while $\Omega_k(b)$ describes the opacity of eigenstate $\phi_k$ from the proton which dissociates and interacts with the same intermediate parton. The normalization constant
\be
A=\pi^2/2g^2_N(0).
\ee
After integrating (\ref{eq:3Rb}) over $t$, the cross section becomes
\be
\frac{M^2 d\sigma_{ik}}{dM^2}~=~A\int\frac{d^2b_2}{\pi}\int\frac{d^2b_1}{2\pi} |\Omega_i(b_2)|^2 \Omega_k(b_1) \cdot S_{ik}^2(\vec{b}_2-\vec{b}_1),
\label{eq:result}
\ee
where here we have included the screening correction $S_{ik}^2$, which depends on the separation in impact parameter space, $(\vec{b}_2-\vec{b}_1)$, of states $\phi_i,\phi_k$ coming from the incoming protons
\be
S_{ik}^2(\vec{b}_2-\vec{b}_1)~\equiv~{\rm exp}(-\Omega_{ik}(\vec{b}_2-\vec{b}_1)).
\ee
If we now account for more complicated multi-pomeron vertices, coupling $m$ to $n$ pomerons, and assume an eikonal form of the vertex with coupling
\be
g^m_n=(g_N\lambda)^{m+n-2},
\label{eq:gmn}
\ee
then we have to replace $\Omega_i$ by the eikonal elastic amplitude and $\Omega_k$ by the inelastic interaction probability.  That is, instead of $\Omega_i(b_2)$ and $\Omega_k(b_1)$, we put
\be
\Omega_i \to 2(1-e^{-\Omega_i(b_2)/2}), ~~~~~~~~~~ \Omega_k \to (1-e^{-\Omega_k(b_1)}).
\label{eq:gmn2}
\ee
Fig. \ref{fig:3Rb}(b) symbolically indicates multi-pomeron couplings.
In (\ref{eq:gmn}), $g_N$ is the proton-pomeron coupling and $\lambda$ determines the strength of the triple-pomeron coupling.\footnote{In comparison with the (\ref{eq:ob},\ref{eq:ot}) expressions the formula for $\Omega_i$ contains an additional factor $\lambda/\pi$, that is we use (\ref{eq:ob}) with $\Omega_i(t)=g_i(t)g_{3P}(t)\exp(\Delta y_i(\alpha_P(t)-1))/\pi=g_i(t)\lambda g_N(0)\exp(B_{3P}t+\Delta y_i(\alpha_P(t)-1))/\pi$ where we assume the exponential dependence of $g_{3P}(t)\propto \exp(B_{3P}t)$ (for the each pomeron leg; see eqs.(4.6) and (4.7) of \cite{Luna}). Here $\Delta y_i$ is the rapidity interval between the proton ($i$) and the triple-pomeron vertex (intermediate parton); $\pi$ in the denominator comes from the definition of the
multi-Reggeon couplings; see an extra $\pi$ ($1/16\pi^2$ and not $1/16\pi$ as in usual elastic cross section) in (\ref{eq:3P}). $t$ dependence of the vertex is parametrized by conventional exponent with the slope $B_{3P}=0.7$ GeV$^{-2}$ for each pomeron leg which is in agreement with the latest H1 data on diffractive $J/\psi$ production with proton dissociation~\cite{H1} and with the results ($B_{3P}<1$/GeV$^2$ is small) of the previous triple-Regge analysis~\cite{KKPT,Luna}.}

\subsection{Implications of TOTEM data for $\SD$ at high mass}
There are indications that the data for high-mass dissociation are not in agreement with the $M$ and $s$ dependence expected from the form of ${M^2 d\sigma}/{dtdM^2}$, based on (\ref{eq:3P}), assuming a {\it constant} $\lambda$.  From (\ref{eq:3P}) we see that the cross section 
should increase with decreasing 
 $M^2$ as 
 \be
 (1/M^2)^{2\alpha_P(t)-\alpha_P(0)-1}~\sim ~(M^2)^{-\Delta}.
 \label{eq:M}
 \ee
 However, the preliminary TOTEM data at $\sqrt s=7$ TeV,  give cross sections integrated over the $3.4 <M< 8$ and $8<M<350$ GeV mass intervals of 1.8 and 3.3 mb respectively \cite{TO4}.  This translates into a cross section ${M^2 d\sigma}/{dtdM^2}$  
 more than twice ($\sim2.4$) smaller for $M$ values in the second as compared to the first mass interval, whereas  (\ref{eq:M}) predicts only about a 60$\%$ increase.
 This observation indicates that the value of $\lambda$ (which specifies the multi-pomeron coupling) should be smaller in the second mass interval.  Secondly, since $\alpha_P(0)>1$, the cross section
for fixed $M^2$ should increase with energy ($\sqrt s$). On the other hand, the TOTEM result is about factor 1.7 less than that measured by CDF at the Tevatron ($\sqrt s=1.8$ TeV). Of course at the higher LHC energy we have a stronger suppression caused by the gap survival factor $S^2_{ik}$ (which was not included in the simplified expression (\ref{eq:3P})), but this is not enough to explain the discrepancy. (Note that the eikonal $S^2$ suppression is rather well fixed after the model was tuned to describe the elastic scattering and low-mass dissociation data.)

So we have phenomenological arguments in favour of introducing some energy dependence of $\lambda$, which specifies the multi-pomeron couplings via (\ref{eq:gmn}).  Since the $g^m_n$ coupling is a dimensionful quantity and the characteristic transverse momenta of the intermediate partons inside the pomeron ladder (i.e. the size of the pomeron) depend on the rapidity of corresponding partons, it looks natural to take 
\be
\lambda\propto 1/k^2_t(y)\ .
\label{eq:lambda1}
\ee
The diffusion in ln$k_t^2$ occurs from both the beam and target sides of the ladder. Following (\ref{eq:D}) we take $k_T^2 \propto (x_0/x)^D$ for diffusion from one side and $k_T^2 \propto (x_0/x')^D$ from the other side; where to evaluate $x'$ we use the relation $xx's=\langle m^2_T\rangle$, and assume $\langle m^2_T\rangle = s_0=1~\GeV^2$.
So we parametrize $k_t(y)$ by
\be
k^2_t=k^2_0
 \left(\left(\frac{x_0}x\right)^D+\left(\frac{x_0}{x'}\right)^D\right)\ ,
\label{eq:lambda2}
\ee 
where we take the same $D=0.28$ and evolve from the same starting point $x_0=0.1$ as (\ref{eq:D}) for $\gamma_i$ of (\ref{eq:gam-i}). We calculate  $x'$ as $x'=s_0/xs$ with $s_0=1$ GeV$^2$.  If $x>x_0$ we replace the $x_0/x$ ratio by 1, and similarly for $x'$.  

After we introduce the dependence of the multi-pomeron couplings on $x$, via 
 (\ref{eq:lambda1}) and (\ref{eq:lambda2}), the values obtained for the single proton 
 dissociation cross section (integrated over the three mass intervals used by TOTEM \cite{TO4}) are shown in Table \ref{tab:1}. 
\begin{table} [h]
\begin{center}
\begin{tabular}{|l|c|c|c|}\hline
 Mass interval (GeV) &   (3.4,~8) & (8,~350)&  (350, 1100)   \\ \hline
  Prelim. TOTEM data & 1.8 & 3.3  & 1.4   \\
 CMS data &  &  4.3 &  \\
 Present model & 2.3  & 4.0 & 1.4 \\
 \hline

\end{tabular}
\end{center}
\caption{\sf The values of the cross section (in mb) for single proton dissociation (integrated over the three    indicated mass intervals) as observed by TOTEM \cite{TO4}, compared with the values obtained in the present model. Recall that TOTEM claims that their preliminary measured cross sections have about 20\% error bars. Note that the value quoted for the CMS \cite{CMSdiff} cross section of dissociation is integrated over the 12 - 394 GeV $M_X$ interval (close to, but in terms of $\ln M_X$, a bit smaller than, the interval (8 - 350) GeV chosen by TOTEM).}
\label{tab:1}
\end{table}
We see that the agreement with the mass dependence of the TOTEM data is now satisfactory. The $t$-slopes, defined by
\be
d\sigma_{\rm SD}/dt~\propto~e^{-B|t|},
\ee
evaluated, using the present model, for the interval $0.02<|t|<0.1~\GeV^2$, for the three mass TOTEM intervals are $B=8.5, ~ 7.2, ~ 6.0 ~\GeV^{-2}$ respectively (the preliminary TOTEM slopes are $B=10.1,~8.5,~6.8~\GeV^{-2}$; in agreement with the theoretical results within the experimental 15\% error bars).

To obtain the model predictions listed in Table \ref{tab:1}, we have included in the last mass interval the contribution of the secondary RRP term using the value of the RRP vertex found in the triple-Regge fit of \cite{Luna}. In the other two mass intervals such a contribution is negligible (less than 0.02 mb). We do not include the PPR contribution since it is dual to the low-mass proton excitations, which in our approach are accounted for in terms of the G-W diffractive eigenstates. 

In the present analysis we have taken $\lambda$ of (\ref{eq:gmn}) to be energy dependent, However, we find $\lambda=0.18$ at relatively low energies when both $x>x_0$ and $x'>x_0$ such that $\lambda$ ceases to be energy dependent. This value is in agreement with the previous triple-Regge analysis of \cite{KKPT,Luna}.

\subsection{Tension between high-mass single dissociation data}

Although TOTEM have made the most detailed observations of high-mass single proton dissociation in high energy $pp$ collisions, the present `global' diffractive  model has been tuned to simultaneously describe the TOTEM data {\it together} with earlier measurements of single dissociation. Here we compare with the description of measurements made by CDF at the Tevatron, \cite{GM} and, later, in Section \ref{sec:5.3}, we show the description of information obtained by ATLAS \cite{atl}.

\begin{figure} 
\begin{center}
\vspace{-6.cm}
\includegraphics[height=14cm]{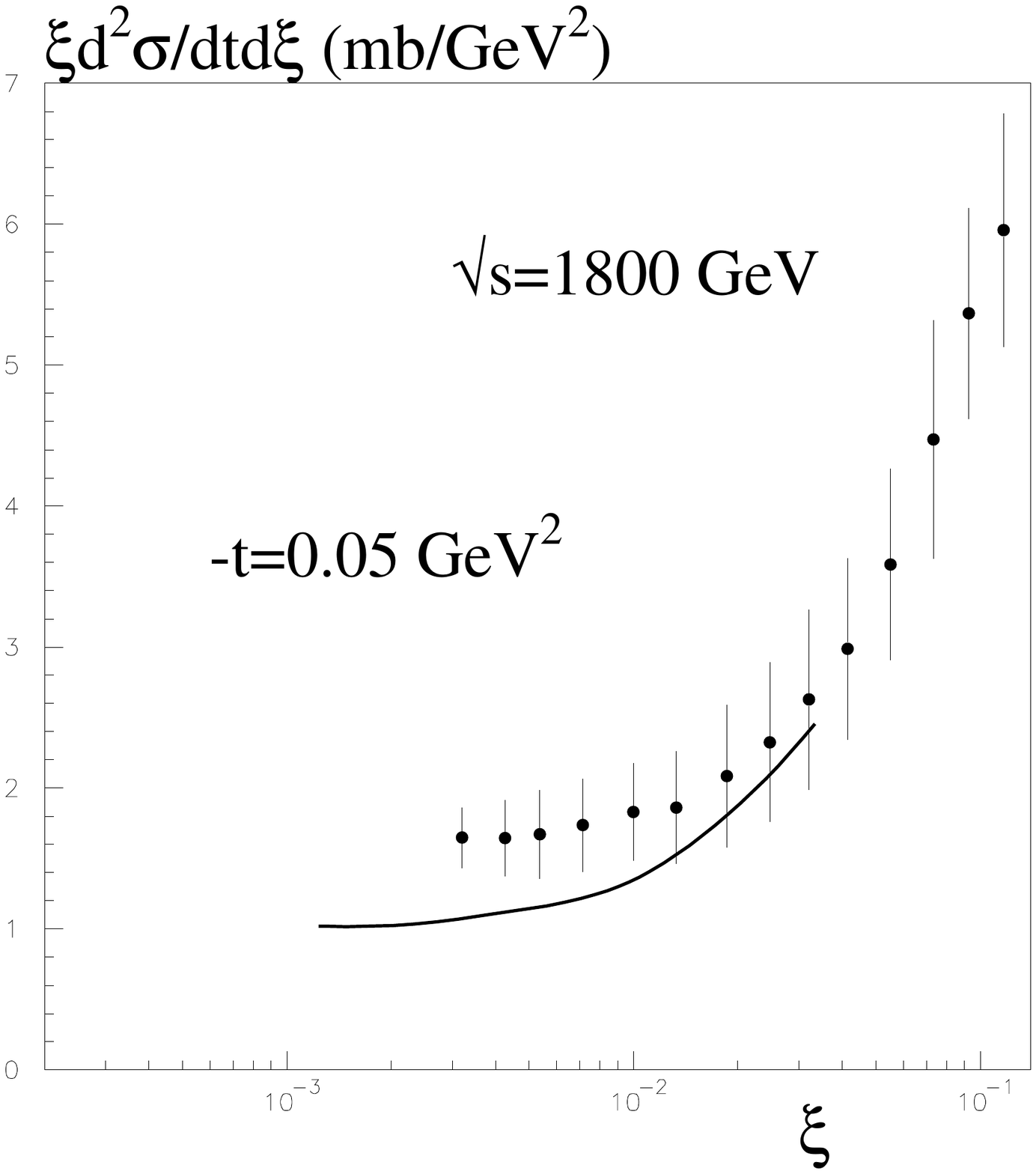}
\vspace*{-.5cm}
\caption{\sf The comparison of the model with data for single proton dissociation measured by the CDF collaboration, given in \cite{GM} but not including a normalisation uncertainty of about 10-15\%. The inclusion of the secondary Reggeon contribution RRP is responsible for the rise of the curve as $\xi$ increases.}
\label{fig:CDF}
\end{center}
\end{figure}

The comparison of the model with the cross section of single proton dissociation observed by the CDF collaboration at $\sqrt{s}=1800$ TeV and $-t=0.05 ~\GeV^2$ is shown in Fig. \ref{fig:CDF}.  We see that the agreement with the CDF data is not particularly good. However, note that: (a) there is some tension between the TOTEM data on the one hand, and CDF results (as well as those of ATLAS and CMS) on the other hand, which enforce us to tune the parameters in such a way that we overestimate the TOTEM single dissociation data, but simultaneously underestimate CDF, ATLAS and CMS cross sections, (b) actually these results were not published by the CDF collaboration, but were published in a separate paper by Goulianos-Montanha \cite{GM} and a normalization uncertainty of about 10 - 15\% was not included in the error bars.

\subsection{Double dissociation and multi-pomeron contributions} 
Recall the puzzle of (\ref{eq:rexp}) and (\ref{eq:rth}); that is the discrepancy,
\be
\frac{\DD~\el}{(\SD)^2}~\simeq~ 3.6~{(\rm expt.)~~versus~~5.0~(naive~theory)},
\ee
between the observation and the prediction of the relation inter-connecting $\el$, $\SD$ and $\DD$.  

Our model already gives satisfactory values for $\el$ and $\SD$.
Below, we therefore consider the possibility that a value of $\DD$ consistent with the TOTEM results can be obtained by the inclusion  of more detailed properties of our present model: the forms of the distributions in $b$-space, the multi-pomeron effects etc.  The multi-pomeron vertices were already included in our description of high-mass single dissociation, see eqs. (\ref{eq:gmn}) and (\ref{eq:gmn2}).

We start with the simplest expression for the double-dissociative cross section, corresponding to the process $pp \to X_1+X_2$ diagram shown in Fig. \ref{fig:3}(a); that is
\be
\frac{M^2_1 M^2_2 d\DD}{dtdM^2_1 dM^2_2}~=~\frac{g^2_{3P}(t)g_N^2(0)}{16\pi^3}~\left(
\frac{M^2_1}{s_0}\frac{M^2_2}{s_0}\right)^{\alpha_P(0)-1}
e^{2|\eta_2-\eta_1|(\alpha_P(t)-1)}\ ,
\label{eq:4DD}
\ee
where we have neglected the survival factor $S^2$. Here $M_1$ and $M_2$ are the masses of the dissociating systems from the two colliding protons, and the $\eta_i$ are the (pseudo)rapidities shown on the diagram. If we now integrate over the square of the momentum transferred, $t$, around the pomeron loop, and express
the opacities as a functions of their impact parameters, then (\ref{eq:4DD})
 takes the form
\be
\frac{d\DD}{d\eta_1d\eta_2}~=~\int dt\frac{d\DD}{d\eta_1d\eta_2 dt}~=~ \frac 1{g^4_N}\int d^2b_1 d^2b_2 d^2b_c
\Omega_{c2}(\Omega_{12}/2)^2 \Omega_{1a}e^{-\Omega_{ac}|\b_a-\b_c|}\ ,
\label{eq:Dop}
\ee
where now we have included the rapidity gap survival factor 
\be
S^2=\exp(-\Omega_{ac}(|\b_a-\b_c|)).
\ee
The notation $(a,~1,~2,~c)$ is specified in the diagram \ref{fig:3}(a). Here the opacities $\Omega_{1a}$ and $\Omega_{c2}$, between the nucleon and the corresponding triple-pomeron vertex, are defined as in (\ref{eq:ob}) and (\ref{eq:ot}), but since the vertex $g_{3P}=\lambda g_N $, the corresponding opacity contains an additional factor $\lambda$. In the same way, the opacity between the two triple-pomeron vertices contains a factor $(\lambda/\pi)^2$; see the footnote below eq.(\ref{eq:gmn2}). In particular, assuming a pure exponential $t$ dependence, this opacity takes the form
\be
\Omega_{12}(b_{12})=g^2_N\frac{\lambda^2}{\pi^2}~e^{2\Delta|\eta_1-\eta_2|}~
\left(\frac{e^{-b^2_{12}/4B_{12}}}{4\pi B_{12}}\right)\ ,
\label{eq;op12}
\ee
where the slope
\be
B_{12}\equiv B_{DD}/2=2B_{3P}+\alpha'_P|\eta_1-\eta_2|.
\ee
The factor $1/g_N^4$ in the denominator of (\ref{eq:Dop}) arises because in our normalization each opacity $\Omega\propto g^2_N$, while cross section (\ref{eq:4DD}) is proportional to $g^4_N$ only.  

To account for the {\em multi}-pomeron vertices, we have to replace $\Omega_{c2}$ and $\Omega_{1a}$ by the inelastic interaction probabilities $(1-\exp(-\Omega_{c2}))$ and $(1-\exp(-\Omega_{1a}))$, while the factor $(\Omega_{12}/2)^2$ is replaced by the probability of elastic parton ``$12$'' scattering, that is by $(1-\exp(-\Omega_{12}/2))^2$.
 Note that, after this eikonal unitarization, we now have no divergency in $\DD$ even in the case of a zero slope $B_{12}$; that is, even for $B_{3P}=0$ and $\alpha'_P=0$. Such a divergency which occurs in (\ref{eq:Dop}), due to the divergency of the $t$ integral and the corresponding divergency of $\Omega_{12}$ for $b_{12}=0$,
is now protected by  the parton ``12'' scattering amplitude, $1-\exp(-\Omega_{12}/2)$.

\begin{figure} 
\begin{center}
\vspace*{-.5cm}
\vspace*{-6.5cm}
\includegraphics[height=16cm]{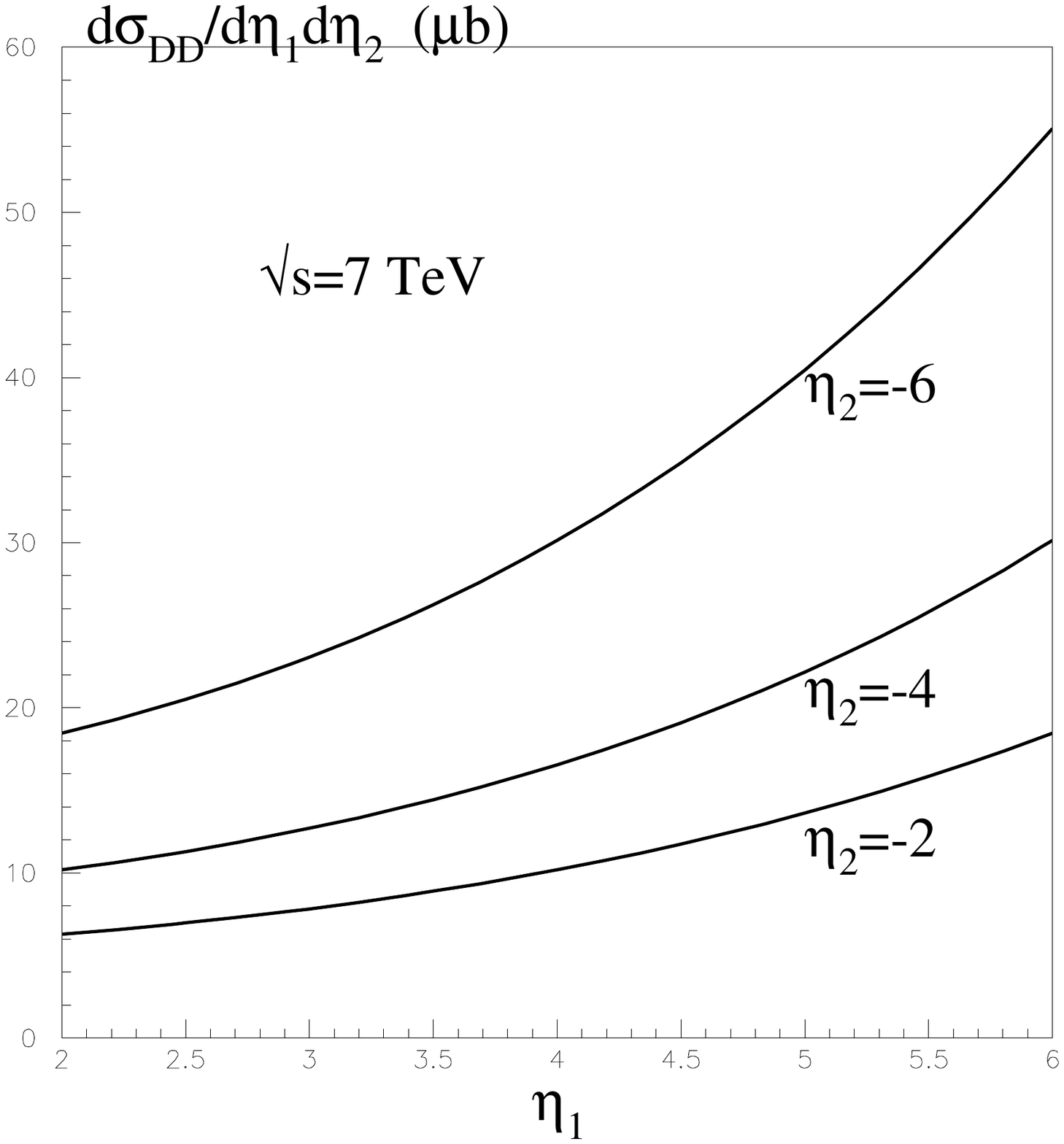}
\vspace*{-.5cm}
\caption{\sf The cross section (in $\mu$b) for double dissociation, $d\DD/d\eta_1 d\eta_2$, at the 7 TeV LHC, as a function of the position of the rapidity gap from $\eta_1$ to $\eta_2$, predicted by the present model which gives a `global' description of high energy elastic and diffractive data.}
\label{fig:C}
\end{center}
\end{figure}

In addition, the {\em multi}-pomeron vertices $g^m_n$ generate  gap survival factors with respect to ``1c'' and ``2a'' inelastic interactions. Overall this gives a screening factor
\be
 \exp(-\Omega_{2a}(|\b_2-\b_a|)-\Omega_{1c}(|\b_1-\b_c|))\ .
\label{s-enh}
\ee 
Thus, finally, we obtain
$$\frac{d\DD}{d\eta_1d\eta_2}~=~ \frac 1{g^4_N}\int d^2b_1 d^2b_2 d^2b_c
(1-e^{\Omega_{c2}})(1-e^{\Omega_{12}/2})^2 (1-e^{\Omega_{1a}})
~~\times$$
\be
~~~~~~~~~~\times~~ {\rm exp}(-\Omega_{ac}(|\b_a-\b_c|)-\Omega_{2a}(|\b_2-\b_a|)-\Omega_{1c}(|\b_1-\b_c|))\ . \label{eq:DDf}
\ee
Typical predictions for the differential cross section of double-dissociation, integrated over the $t$, are shown in Fig. \ref{fig:C}. They correspond to our `global' description of diffractive data, and account for the $k_t$ dependence of $\lambda$, keeping all the parameters determined as described in the previous Sections.

After the integration over the $-4.7~>\eta_2>-6.5$ and $4.7<\eta_1<6.5$ rapidity intervals covered by TOTEM, we obtain $\DD=145 \ \mu$b, close to the upper bound of the TOTEM measurement $116\pm 25\ \mu$b \cite{TO3}.   It is encouraging that the more physical and complicated structure of the present model largely reconciles the discrepancy between
(\ref{eq:rexp}) and (\ref{eq:rth}).

\subsection{Large Rapidity Gaps in central region, and SD and DD \label{sec:5.3}}

\begin{figure} 
\begin{center}
\vspace*{-6.50cm}
\includegraphics[height=15cm]{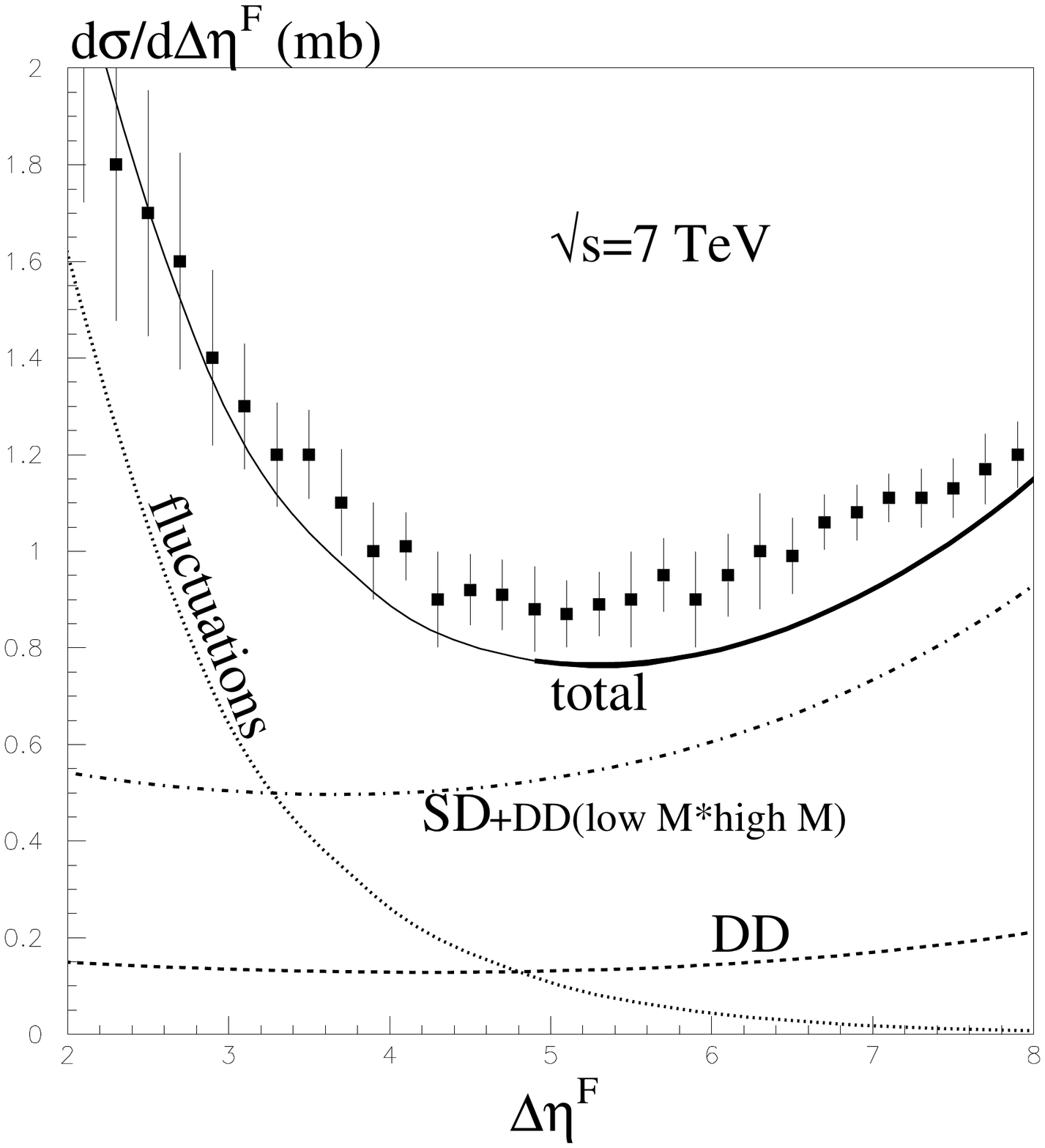}
\vspace{-0.0cm}
\caption{\sf The ATLAS \cite{atl} measurements of the inelastic cross section differential in rapidity gap size $\Delta\eta^F$ for particles with $p_T>200$ MeV. Events with small gap size ($\Delta\eta^F \lapproxeq 5$) may have a non-diffractive component which arises from fluctuations in the hadronization process \cite{FZ}. This component increases as $\Delta\eta^F$ decreases (or if a larger $p_T$ cut is used \cite{FZ,atl}).  
The data with $\Delta\eta^F\gapproxeq 5$ are dominantly of diffractive origin, are compared with the present `global' diffractive model.}
\label{fig:eta}
\end{center}
\end{figure}

The ATLAS \cite{atl} and CMS \cite{CMSdiff} collaborations have measured the cross section of events with a large rapidity gap, $\Delta\eta^F$, which starts before the edge of the forward calorimeter
($\eta=4.9$ for ATLAS) and ends somewhere inside the opposite forward  calorimeter or in the tracking central detector. The ATLAS data are shown in Fig. \ref{fig:eta}, and correspond to measurements of the inelastic cross section differential in the size of the rapidity gap $\Delta\eta^F$ for particles with $p_T>200$ MeV. When $\Delta\eta^F$ decreases below about 5, the data are increasingly contaminated by fluctuations from the hadronisation process, but for $\Delta\eta^F \gapproxeq 5$ they are a measure of proton dissociation; in fact 
mainly of single proton dissociation. That is, the LRG actually starts just from a leading proton. However, we should not neglect the contribution of events where both protons dissociate, but the secondaries produced by one proton, say, the $M_X$-group, go into the beam pipe and are not seen in the calorimeter. In Fig. \ref{fig:eta} this double dissociation contribution is shown by the dashed curve.
\newpage

It was demonstrated in \cite{FZ} that, depending on the particular mechanism of hadronization, the fluctuations may be able to account for the data at rather small $\Delta\eta^F$. To allow phenomenologically for such a possibility we  assume an exponential dependence
of this contribution, $\propto \exp(-a|\Delta\eta|)$ with $a=0.9$. If this term is normalized to the ATLAS data \cite{atl} then it gives the dotted line in Fig. \ref{fig:eta}.

\section{Discussion  \label{sec:6}}

The high energy diffractive data that are presently available cover a wide variety of processes. These include measurements of  the total and elastic $pp$ cross sections ($\sigma_{\rm tot},~ \el$), the elastic differential cross section $(d\el/dt$), 
the cross sections of low- and high-mass proton dissociation ($\SD^{{\rm low}M},~\SD^{{\rm high}M}$), the cross section of events where both protons dissociate $(\DD^{{\rm high}M}$), as well as the probability of inelastic events with a large rapidity gap ($d\sigma/d\Delta\eta$).

\subsection{$k_t(y)$ effect}   
Here, we demonstrate that all these diffractive data may simultaneously be described within the Regge Field Theoretic framework based on only one pomeron pole. However, to reach  agreement with the data, we have to include pomeron-pomeron interactions, arising from multi-pomeron vertices, and to allow for the $k_t(y)$ dependence of the multi-pomeron vertices. Recall that, due to the BFKL-type diffusion in $\ln k^2_t$ space,  together with the stronger absorption of low $k_t$ partons, the typical transverse momentum, $k_t$, increases with energy depending on the rapidity
 position of the intermediate parton or the multi-pomeron vertex. This $k_t(y)$ effect enables the model to achieve a relatively low probability of low-mass 
 dissociation of an incoming proton and to reduce the cross section of high-mass  
 dissociation in the central rapidity region in comparison with that observed closer 
 to the edge of available rapidity space -- both of which are features demanded by the recent TOTEM data.

Recall that such a non-trivial behaviour of the diffractive cross sections, caused by the energy and rapidity dependence of the $k_t$'s along the ladder, was predicted in \cite{KMR-s3}, see, in particular, Fig. 9 of that paper. Since the growth of the transverse momenta $k_t(y)$ (which leads to a rapidity dependence of the dimensionful multi-pomeron couplings $g^m_n$ and of the proton-pomeron couplings $g_i$ as in (\ref{eq:gamma}) and (\ref{eq:gam-i})) is a crucial ingredient of our model, it would be interesting to observe it directly. This can be done by studying experimentally the transverse momentum distributions of secondaries near the edge of a large rapidity gap as a function of the rapidity position of the gap edge.
In particular, it could be achieved by combining simultaneous measurements in the  TOTEM and CMS detectors, when the rapidity gap is fixed by observing the leading proton in the TOTEM roman pots, while the secondaries produced in the `pomeron'-proton collision are detected by CMS.

\subsection{Notes on describing dissociation data at the LHC}
Even though including the $k_t(y)$ dependence considerably improves the description of the dissociation data, the overall agreement with these data is not particularly good. This is mainly due to a {\em tension} between the TOTEM and the ATLAS, CMS, CDF results\footnote{Such a tension was also emphasized by S. Ostapchenko \cite{Ost1}.}. It is possible to improve the description of the TOTEM data on proton dissociation. We simply need a reduction of about 10 - 15\% of the starting value of $\lambda$, the parameter which specifies the muti-pomeron coupling. However, if we do this, we will even further underestimate the $M^2d\sigma/dM^2$  cross section at the Tevatron, and also the probability to have a LRG in the central rapidity region observed by the ATLAS and CMS\footnote{Recall that the CMS \cite{CMSdiff} cross section of dissociation integrated over the 12 - 394 GeV $M_X$ interval (close to, but in terms of $\ln M_X$, a bit smaller than, the interval (8 - 350) GeV chosen by TOTEM) is noticeably larger (4.3 mb) than that (3.3 mb) found by TOTEM.} groups. 
Here, we have tuned the model to give a compromise solution somewhere  between the CDF (ATLAS/CMS) and the TOTEM results.

It is also possible to obtain a lower value of $\SD$  integrated over the central of the three mass intervals used by TOTEM (while keeping the same cross sections in the low and large $M_X$ intervals)  by choosing a larger value of  the parameter $D$. However, if we were to do this then we would find that the probability of low-mass dissociation, $\sigma_D^{{\rm low }M}$, is too small (due to the small $\langle T^2\rangle -\langle T\rangle ^2$ dispersion caused by $\gamma_{1,2}\to 1$). Moreover, the model would then give an even steeper $d\sigma/d\Delta\eta^F$ behaviour of the LRG cross sections with increasing $\Delta\eta^F$. The model already has $d\sigma/d\Delta\eta^F$ growing a bit faster than the ATLAS and CMS data.

Here, we have adjusted the parameters of the model to give a reasonable description of all aspects of the available diffractive data.  If, instead, we had performed a $\chi^2$ fit to the  data, then the few dissociation measurements of TOTEM (values of $\SD$ in three mass intervals with 20\% errors, and one value of $\DD$) would have carried little weight.
On the  other hand, all the TOTEM data are self-consistent between themselves. Moreover, these data  reveal a very reasonable tendency of the $d\SD/d\xi$ dependence, close to that predicted in \cite{KMR-s3} where the $k_t$ distribution of the intermediate partons inside the pomeron ladder, and the role of the transverse size of the different QCD pomeron components, were accounted for  more precisely.
 Therefore, we have presented the results of this `compromised' description (and not made a $\chi^2$ 
 fit) in order not to discard the interesting new information coming from 
the recent TOTEM measurements\footnote{We do not include in the present description the secondary Reggeon PPR contribution which is partly `dual' to that arising from the G-W diffractive eigenstates. In general, it should be considered in  future `global' diffractive analyses, but at present it does not change the situation qualitatively. So we prefer not to introduce the extra parameters.}.

Of course, it would be best to implement in the model also the effects caused by more complicated (non-local) structure of the original pomeron. At present we consider just one `effective' pomeron pole renormalized by enhanced absorptive corrections. On the other hand, bearing in mind the tension we have seen between the results of the different collaborations, it is too soon to undertake a complete precise analysis, while our present simplified model appears sufficient to describe the qualitative features of the high-energy diffractive data.

\subsection{Predictions of high-energy diffraction observables}
For completeness, we give in Table \ref{tab:2} the values of some of the diffractive observables obtained from the present `global' description of diffractive high energy data.  We include, in particular, the values at collider energies relevant to experiments at the LHC.
\begin{table} [h]
\begin{center}
\begin{tabular}{|r|r|c|c|c|c|c|l|c|c|}\hline
$\sqrt{s}$ & $\sigma_{\rm tot}$ &   $\el$   &  $B_{\rm el}(0)$ & $\SD^{{\rm low}M}$ &     $\DD^{{\rm low}M}$ & $\SD^{\Delta\eta_1}$ &  $\SD^{\Delta\eta_2}$ &  $\SD^{\Delta\eta_3}$ & $\DD^{\Delta\eta}$ \\ \hline
  (TeV)&  (mb)  &    (mb)  & ($\GeV^{-2}$) & (mb) & (mb) & (mb) &(mb)&(mb)&($\mu$b)\\ \hline
    1.8 &  77.0  &   17.4  &   16.8  &  3.4  &   0.2 &  &  &  &   \\
    7.0 &  98.7  &    24.9 &   19.7  &  3.6  &   0.2 & 2.3 & 4.0  & 1.4 & 145 \\
    8.0 &  101.3  &   25.8 &   20.1  &  3.6  &   0.2 & 2.2 & $3.95$ & 1.4 & 139 \\
   13.0 &  111.1  &   29.5 &   21.4 &  3.5  &   0.2  & 2.1 & 3.8 & 1.3 & 118 \\  
   14.0 &  112.7  &   30.1 &   21.6 &  3.5  &   0.2  & 2.1 & 3.8 & 1.3 & 115 \\
  100.0 &  166.3  &   51.5 &   29.4 &  2.7  &   0.1  &  &  &  &  \\
\hline
\end{tabular}
\end{center}
\caption{\sf The predictions of the present model for some diffractive observables for high energy $pp$ collisions at $\sqrt{s}$ c.m. energy. $B_{\rm el}(0)$ is the slope of the elastic cross section at $t=0$. Here $\SD$ is the sum of the single dissociative cross section of both protons. The last four columns are the model predictions for the cross sections for high-mass dissociation in the rapidity intervals used by TOTEM at $\sqrt{s}$=7 TeV: that is, $\SD$ for the intervals $\Delta\eta_1 = (-6.5,-4.7)$, $\Delta\eta_2 = (-4.7,~4.7)$, $\Delta\eta_3 = (4.7,~6.5)$, and $\DD^{\Delta\eta}$ is the double dissociation cross section
where the secondaries from the proton dissociations are detected in the rapidity intervals $\Delta\eta_1 = (-6.5,-4.7)$ and $\Delta\eta_3 = (4.7,~6.5)$,  At $\sqrt{s}$=7 TeV, the three `SD' rapidity intervals correspond, respectively, to single proton dissociation in the mass intervals $\Delta M_1 = (3.4,8)$ GeV, $\Delta M_2 = (8,350)$ GeV, $\Delta M_3 = (0.35,1.1)$ TeV, see Table \ref{tab:1}.}
\label{tab:2}
\end{table}

Recall that the slow rise of $\SD^{{\rm low}M}$  from a model value of 2.6 mb at the CERN-ISR energy to the value 3.6 mb at the LHC energy of $\sqrt{s}$=7 TeV is due to the growth of the characteristic momentum of the pomeron, $k_t^2 \propto s^D$, see (\ref{eq:D}). We noted that this behaviour agrees with the TOTEM measurement of low-mass dissociation \cite{TO2}. Also, as just mentioned above, the energy dependence of the characteristic $k_t$ of the pomeron, which translates into a rapidity dependence, $k_t(y)$, is  in accord with the preliminary TOTEM measurements of $\SD$ in the three different mass (or rapidity) intervals, see Table \ref{tab:1}. The decrease of the cross sections for dissociation at $\sqrt{s}$=100 TeV, seen in Table \ref{tab:2}, is because we are beginning to approach the true black disk limit, where the probability of dissociation tends to zero,  while the effective
$\alpha'_{\rm eff}=\frac 12 dB_{\rm el}/d\ln s$ of elastic slope  increases.  

The values listed in Table \ref{tab:2} for $\sqrt{s}$=7 TeV are highly constrained by the recent LHC Run 1 measurements. These measurements therefore largely determine the high energy predictions of the model. When more precise and extensive diffractive data become available, and the tensions between data sets are reduced, the model predictions may have to be adjusted.

\section*{Appendix A: Unified `soft' and `hard' pomeron}
`Soft' and `hard' high-energy $pp$ interactions are described in different ways.  As we have discussed, the appropriate formalism for high-energy soft interactions is based on Reggeon Field Theory with a phenomenological (soft) pomeron, whereas, on the other hand, for hard interactions a QCD partonic approach is used, where the (QCD) pomeron is associated with the BFKL vacuum singularity \cite{book}. However, the two approaches appear to merge naturally into one another.  That is, the partonic approach seems to extend smoothly into the soft domain. 

\begin{figure}[htb]
\begin{center}
\includegraphics[height=10cm]{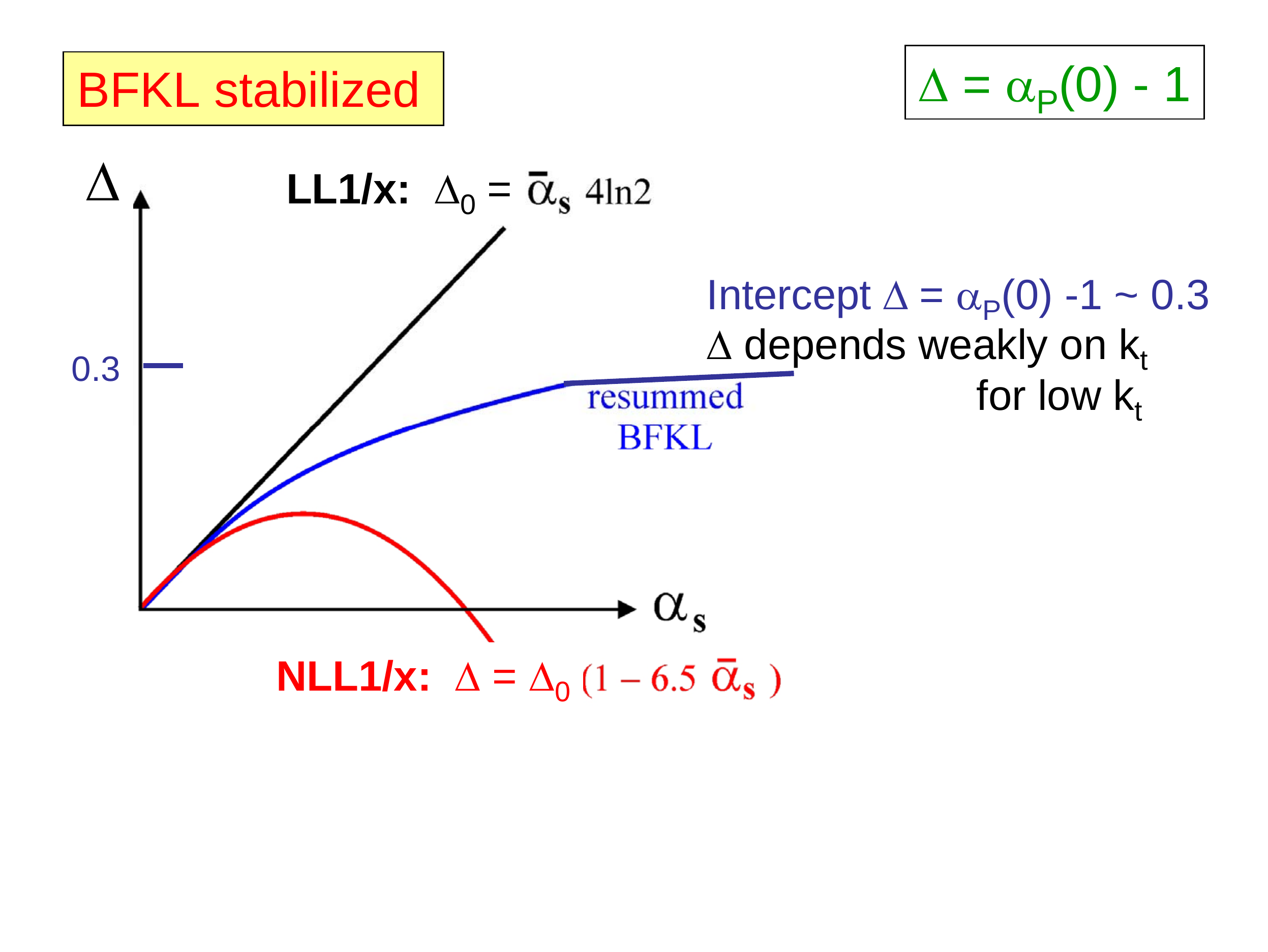}\hspace{1cm}
\vspace{-3cm}
\caption{\sf The behaviour found for the pomeron intercept at leading and next-leading log$(1/x)$ order, where $\bar{\alpha}_s\equiv\alpha_s/3\pi$. When an all-order resummation of the main high-order contributions is included, $\Delta$ tends to a value of about 0.3 for reasonably large values of $\alpha_s$.}
\label{fig:BFKLstab}
\end{center}
\end{figure}

The BFKL equation describes the development of the gluon shower as the momentum fraction, $x$, of the proton carried by the gluon decreases.  That is, the evolution parameter is ln$(1/x)$, rather than the ln$k_t^2$ evolution of the DGLAP equation. 
Formally, to justify the use of perturbative QCD, the BFKL equation should be written for gluons with sufficiently large $k_t$. However, it turns out that, after accounting for NLL$(1/x)$ corrections and performing an all-order resummation of the main higher-order contributions \cite{resum1,resum2,resum3}, the intercept of the BFKL pomeron depends only weakly on the scale for reasonably small scales. The intercept is seen to be $\Delta \equiv \alpha_P(0)-1 \sim 0.3$ over a large interval of smallish $k_t$, Fig.~\ref{fig:BFKLstab}.
Thus the BFKL pomeron is a natural object to continue from the `hard' domain into the `soft' region.
\begin{figure}[htb]
\begin{center}
\vspace{-1.1cm}
\includegraphics[height=8cm]{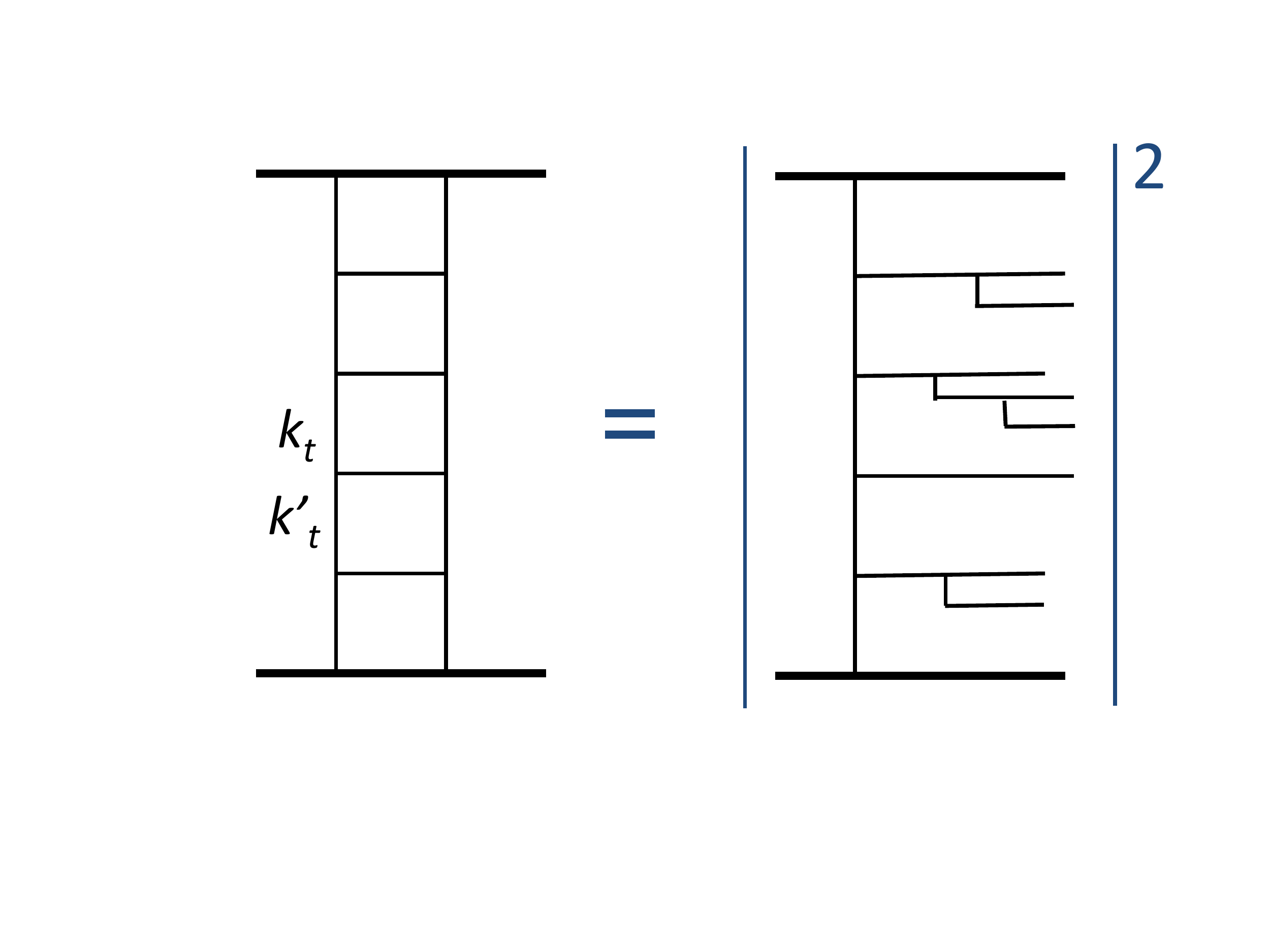}
\vspace{-2.cm}
\caption{\sf The cascade structure of a gluon ladder. The BFKL or QCD pomeron is the sum of ladder diagrams, each with a different number of rungs.}
\label{fig:cascade}
\end{center}
\end{figure}

The BFKL or QCD pomeron may be viewed as a sum of ladders based on the exchange of two $t$-channel (Reggeized) gluons. Each ladder produces a gluon cascade which develops in ln$(1/x)$ space, and which is not strongly ordered in $k_t$, see Fig.~\ref{fig:cascade}. 

Let us discuss the behaviour of the $k_t$'s along the ladder in more detail. For proton-proton interactions we have the same initial relatively low scale at both ends of the ladder. So there is no reason (or direction) for a strong ordering of $k_t$'s along the ladder. In the original Regge theory it was assumed that the values of the $k_t$'s were limited.
Indeed, consider the simplest $\lambda \phi^3$ theory with a dimensionful coupling $\lambda$ measured in GeV. This theory is superconvergent. The $k_t$ integral corresponding to one cell of the ladder of Fig. \ref{fig:cascade} is of the form
\be 
\lambda^2\int \frac{d^2k_t}{(k_t^2+m^2)^2},
\ee
where the typical value of $k_t\sim m$ is driven by the mass of the $\phi$ meson.  Nowadays\footnote{A more detailed discussion of this historical development is given in \cite{KMRJphysG}.}, with QCD, we have a logarithmic theory with a dimensionless coupling $\alpha_s(k^2)$. The corresponding integral takes the form
\be 
\int \frac{\alpha_s(k_t^2)d^2k_t}{k_t^2+k_t^{'2}},
\ee
where now the value of $k_t$ is determined by the value of $k'_t$ of the $t$-channel gluon in the neighbouring cell. Due to the up/down symmetry at each evolution step, the $k_t$ of the gluon may become a few times larger or smaller with equal probabilities. In this way we obtain the famous BFKL diffusion in ln$k_t$ space \cite{Lipatov}. Accounting for the running $\alpha_s(k_t^2)$ shifts the `diffusion' in the direction of lower $k_t$. On the other hand, the low $k_t$ gluon has a much larger probability to be absorbed, $\sigma_{\rm abs} \sim 1/k^2_t$.  Therefore, finally, the absorptive corrections described by the enhanced diagrams (which were not included explicitly in the present model\footnote{The enhanced absorptive corrections, which depend on $k_t$, were included and studied in \cite{KMR-s3} where, indeed, we observed this effect; that is, the increase of $k_t$ during the BFKL evolution.}) push the BFKL diffusion to enlarge the $k_t(y)$.

There are phenomenological arguments (such as the small slope of the pomeron trajectory\footnote{Recall that $\alpha'_P \propto 1/\langle k_t^2 \rangle \propto R_{\rm Pom}^2$.}, the success of the Additive Quark Model relations\footnote{The argument is best seen by analogy with nuclear physics. For light nuclei we have  `additive' cross sections, $\sigma=A_1A_2 \sigma_{NN}$, since the nuclei radii $R \gg r_{NN}$ and the nucleons do not screen each other,
where $\sigma_{NN}$ and $r_{NN}$ are the nucleon-nucleon cross section and interaction radius. On the other hand, for a heavy nucleus large Glauber corrections break the additive result. Similarly, the experimental success of the AQM indicates that $r_{qq}\sim R_{\rm Pom} \ll R_p$ -- the proton radius.}, etc.) which indicate that the size of an individual pomeron is relatively small as compared to the size of a proton or pion etc. Thus we may regard the cascade as a small-size `hot-spot' inside the colliding protons.

At LHC energies the interval of BFKL ln$(1/x)$ evolution is much larger than that for DGLAP ln$k_t^2$ evolution. Of course, it is not enough to have only one pomeron ladder exchanged; we need to include multi-pomeron exchanges -- that is we must perform eikonal unitarisation (so-called absorptive corrections) as discussed above.

\begin{figure}[htb]
\begin{center}
\vspace{-.0cm}
\includegraphics[height=10cm]{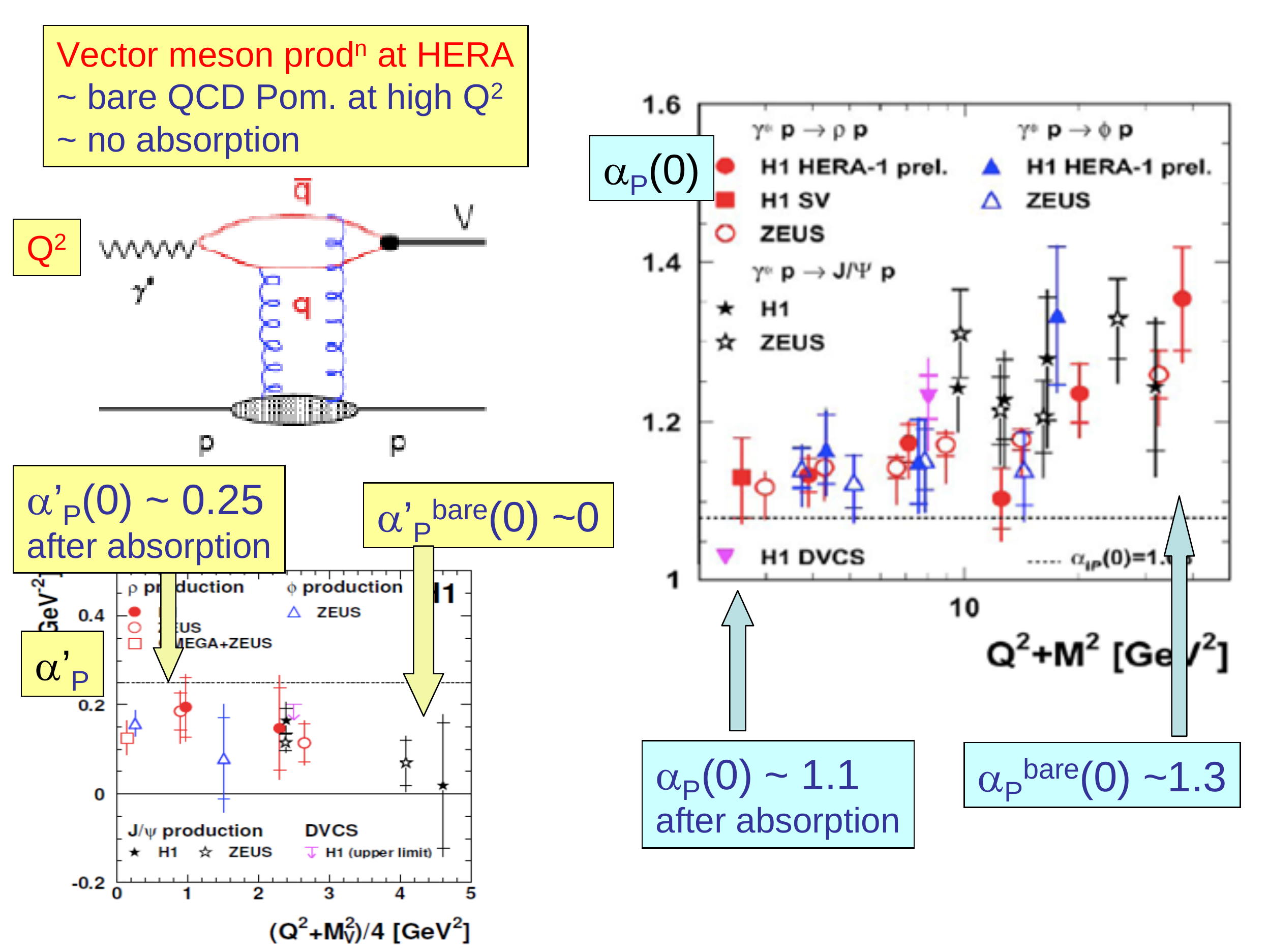}\hspace{1cm}
\caption{\sf The parameters of the pomeron trajectory, $\alpha_P(t)=\alpha_P(0)+\alpha'_Pt$, determined from the energy and $t$ behaviour of high energy HERA data for vector meson production, $\gamma^*p\to V(M)p$.}
\label{fig:VM}
\end{center}
\end{figure}

Basically, the picture is as follows. In the perturbative domain we have a single bare `hard' pomeron exchanged with a trajectory $\alpha_P^{\rm bare}\simeq 1.3+\alpha'_{\rm bare}t$, where $\alpha'_{\rm bare} \lapproxeq 0.05$ GeV$^{-2}$. The transition to the soft region is accompanied by absorptive multi-pomeron effects, such that an {\it effective} `soft' pomeron may be approximated by a linear trajectory $\alpha^{\rm eff}_P \simeq 1.08+0.25t$ in the {\it limited} energy range up to Tevatron energies \cite{DL}.  This smooth transition from hard to soft  is well illustrated by Fig.~\ref{fig:VM}, which shows 
 the behaviour of the data for vector meson ($V=\rho, \omega, \phi, J/\psi$) production at HERA, $\gamma^*p\to V(M)p$, as $Q^2+M^2$ decreases from about 50 GeV$^2$ towards zero\footnote{Note, however, that only part of the $Q^2$ dependence of the effective intercept in Fig.~\ref{fig:VM} is due to absorptive corrections. Another part is due to the double log dependence of the deep inelastic scattering amplitude, $A(x,Q^2)\propto {\rm exp}(\sqrt{(4N_c\alpha_s/\pi)~{\rm ln}(1/x)~{\rm ln}Q^2})$.}.

\section*{Appendix B: Low-Nussinov pomeron as an example of $\sigma_{\rm abs}\sim 1/k_t^2$}
At high energies a good example of a G-W eigenstate is a state formed by valence quarks, whose position in the impact parameter ($b$) plane is fixed.  In other words, the interaction with the QCD pomeron (that is, with two $t$-channel gluons) does not change the $b$ coordinates. Consider simple two-gluon Low-Nussinov \cite{Low,Nus} pomeron exchange.  In this case, the cross section for pomeron exchange between two quark dipoles is given by
\be
\sigma_{ab}~=~\frac{32\pi}{9}\int \frac{dk_t^2}{k_t^4}~ \alpha^2_s~[1-F_a(4k_t^2)]~[1-F_b(4k_t^2)].
\label{eq:71}
\ee 
Here the infrared divergency at small $k_t$ is cutoff by the interaction with the quark spectators. In the simplified dipole model this effect is described by the factors [...] in the numerator, where $F_i(4k^2_t)$ are the form factors of the incoming colourless dipoles. Due to this cutoff, the cross section $\sigma_{ab} \propto \alpha^2_sr^2$. That is, a larger size G-W  component corresponding to a  larger $r$, has a larger cross section. In other words, the cross section $\sigma_{ab}\propto 1/k^2_{\rm min}$, where $k_{\rm min}$ provides the effective infrared cutoff for the convergency of the integral in (\ref{eq:71}). Here, the convergency is provided by the interaction with the quark spectator. Note that in this simple model $\sigma_{ab}$ of (\ref{eq:71}) is the inelastic, that is the absorption, cross section.

If, on the contrary, the integral is cutoff at a larger $k_t$ by some $k_{\rm min}$ arising from the internal structure of the effective pomeron (in a region where the $F_i(4k^2_t)\ll 1)$, then the cross sections of the different G-W components will be practically the same.

That is, the value of the cross section is specified by the cutoff induced by the pomeron, and not by the size of the G-W eigenstates. As a consequence, all eigenstates have the same cross section, so there is no dispersion, and the interaction will not destroy the coherence of  the wave functions of the incoming protons. Hence, the probability of diffractive dissociation will be negligible. As was discussed in \cite{KMRJphysG}, the value of $k_{\rm min}$ increases with energy. This behaviour was shown theoretically in \cite{Ryskin:2009tj,KMR-s3}, and phenomenologically it was observed in the tuning of the Pythia8 Monte Carlo \cite{P8}, where the cutoff has the behaviour
\be
k^2_{\rm min}~\sim ~s^{\beta} ~~~~~~ {\rm with} ~~~~~~~~\beta\simeq 0.24.
\ee

\section*{Acknowledgements}

We thank Kenneth Osterberg, Paul Newman and Sergey Ostapchenko for discussions, and to Christophe Royon for urging us to write this article.   MGR thanks the IPPP at the University of Durham for hospitality. This work was supported  by the Federal Program of the Russian State RSGSS-4801.2012.2.

\bibliographystyle{JHEP.bst}
\bibliography{F}

\end{document}